\documentclass{JHEP}
\usepackage{graphicx}
\usepackage{epsfig}
\usepackage{amssymb}
\usepackage{amsmath}
\usepackage{float}

\newcommand{\bea}{\begin{eqnarray}}
\newcommand{\eea}{\end{eqnarray}}
\newcommand{\bean}{\begin{eqnarray*}}
\newcommand{\eean}{\end{eqnarray*}}
\newcommand{\nn}{\nonumber \\}

\def\O #1{\overline{#1}}

\def\W #1{\widetilde{#1}}
\def\WH #1{\widehat{#1}}

\def\braket#1{\left\langle #1 \right\rangle}

\def\ket#1{\left| #1\right\rangle}
\def\gb #1{ \left\langle #1 \right]}
\def\tgb #1{ \left[ #1 \right\rangle}

\def\wtl{\widetilde{\lambda}}
\def\wh{\widehat}

\def\a{{\alpha}}

\def\b{{\beta}}

\def\la{\lambda}

\def\vev{\braket}
\def\tgb #1{ \left[ #1 \right\rangle}
\def\bket#1{\left| #1\right]}
\def\bvev#1{\left[ #1 \right]}
\def\Spaa{\vev}
\def\Spbb{\bvev}
\def\Spab{\gb}
\def\Spba{\tgb}

\def\Label#1{\label{#1}%
  \smash{\hbox to0pt{\raise1ex\hbox{\tiny[#1]}\hss}}}

\title{BCFW Recursion Relation with Nonzero Boundary Contribution}
\author{Bo Feng$^{a}$\footnote{email address: b.feng@cms.zju.edu.cn},
 {Junqi Wang$^{b}$}, {Yihong Wang$^{b}$},
{Zhibai Zhang$^{c}$}\\$^{a}${Center of Mathematical Science,
Zhejiang
 University,Hangzhou, China}\\$^{b}${Physics Department, Zhejiang
 University, Hangzhou, China}\\$^{c}${Mathematics Department, Zhejiang
 University, Hangzhou, China}}

\date{\today}
\abstract{The appearance of BCFW on-shell recursion relation has
deepen our understanding of quantum field theory, especially the one
with gauge boson and graviton. To be able to write the BCFW
recursion relation, the knowledge of boundary contributions is
needed. So far, most applications have been constrained to the cases
where the boundary contribution is zero.  In this paper, we show
that for some theories, although there is no proper deformation to
annihilate the boundary contribution, its effects can be analyzed in
simple way, thus we do able to write down  the BCFW recursion
relation with boundary contributions. The examples we will present
in this paper include the $\la\phi^4$ theory and Yukawa coupling
between fermions and scalars. }

\begin{document}

\newpage

\section{Motivations}

For theory with Lagrangian description, we can calculate amplitudes
using  Feynman diagrams. Any Feynman diagram is constructed by
putting some elements, i.e., the vertex, together through
propagators. Thus any higher point amplitude can be constructed
recursively from lower point amplitudes with one very important
feature: {\sl these lower point amplitudes must be well-defined
off-shell}. Comparing to the on-shell amplitudes, which have
physical meaning, off-shell amplitudes are usually longer and more
complicated, especially for gauge theory where gauge freedom renders
the expression including many redundent information. Thus it is
natural to ask if we can construct any higher point on-shell
amplitude recursively from lower point on-shell amplitudes only. If
we could, we can call this theory "on-shell constructible" to be
distinguish with off-shell constructions by Feynman diagrams.

Initially by Witten's twistor program \cite{Witten:2003nn}, BCFW
recursion relation \cite{Britto:2004ap, Britto:2005fq} provides the
first concrete example for on-shell constructibility. Let us first
review how the goal is achieved. First we pick up  two special
momenta $p_1, p_2$ and do the following deformation (BCFW
deformation) using an auxiliary momentum $q$:
\bea p_1(z)=p_1+z q,~~~~p_2(z)=p_2-z q~.
~~~~~\label{BCFW-deform-1}\eea
The opposite sign makes the momentum conservation  satisfied.
Furthermore, if we impose the conditions $q^2=0$, $p_1\cdot
q=p_2\cdot q=0$, the on-shell conditions of $p_1(z)$ and $p_2(z)$
are also satisfied. In another word, we have a deformed on-shell
amplitude $A(z)$ over single complex variable $z$. Having the
deformed $A(z)$ we consider following contour integration
\bea B =\oint_C {A(z)\over z}d z\eea
where contour $C$ is big enough circle around $z=0$. We can evaluate
the integration by two different ways: either by contour around
$z=\infty$ or the big contour around the origin. Thus we have
\bea A(z=0)= -\sum_{z_\a} {\rm Res}\left( {A(z)\over
z}\right)+B~,~~~\label{BCFW-rel-1}\eea
where $A(z=0)$ is the amplitude we want to find and $B$ is the
boundary contribution. The residue part can always be calculated
using the factorization properties from lower-point on-shell
amplitudes. In another word, the expression (\ref{BCFW-rel-1}) tells
us that for any theory, some parts of tree amplitudes are "on-shell
constructible". The trouble part comes from the boundary
contribution $B$. It is easy to see that $B\neq 0$ when and only
when $A(z)$ is not zero under the limit $z\to \infty$. Thus if there
is some deformation such that $A(z)\to 0$ when $z\to \infty$, we
will get the wanted on-shell constructibility. Assuming this strong
condition, i.e., $B=0$, some beautiful results can be derived in
\cite{Paolo:2007}.

From above discussions, we see that for the application of on-shell
constructibility, the knowledge of boundary contribution becomes
very important. The analysis of the boundary behavior is, in
general, an extremely nontrivial task for many theories, especially
the one with gauge symmetry, as demonstrated in the beautiful paper
\cite{ArkaniHamed:2008yf} as well as others, for examples
\cite{Vaman:2005dt}-\cite{Cheung:2008dn}. In these papers, it is
shown that for gauge theory or gravity theory, with proper choice of
BCFW deformation we can make the boundary contribution zero and
derive the on-shell recursion relation.

However, there are other theories where we can not find any
deformation to set boundary contribution zero. One typical example
is the $\la\phi^4$ theory. For these theories, the on-shell
constructibility is not so easy to answer. In fact, more accurate
statement from expression (\ref{BCFW-rel-1}) is following: {\sl If
$B=0$, the theory is on-shell constructible, but if $B\neq 0$, it
can be on-shell constructible or not on-shell constructible}.

In this paper, we will address the problem carefully with  $B\neq
0$. We will show that for some theories, although there is no any
choice to set boundary contribution zero, the boundary contribution
can be analyzed and obtained in fairly simple way through lower
point on-shell amplitudes. Thus for these theories, we can still
write down the on-shell recursion relations.

The structure of our paper is following. In section 2, we use the
$\la\phi^4$ theory as our first example to demonstrate the on-shell
constructibility with nonzero boundary contributions.  In section
2.1, we have identified boundary contributions from Feynman diagram
analysis and written down the BCFW recursion relation. Then in
section 2.2, we calculated several amplitudes using our BCFW formula
and compared with results from Feynman diagrams in Appendix A.

There is another way to deal with $\la\phi^4$ theory by introducing
a massive field as given in \cite{Paolo:2007}. For the new
Lagrangian we have a triple deformation with vanishing  boundary
contributions and similar BCFW recursion relation. In section 3, we
use same triple deformation for $\la\phi^4$ theory. We showed that
how the boundary contributions for $\la\phi^4$ theory are mapped to
the pole contributions in the new Lagrangian, thus established the
equivalent  relation between these two methods.

In section 4, we discuss the scalar QCD, i.e., fermion  interacts
with scalar through the Yukawa coupling. This example is more
interesting because this kind of interactions is a major part of
standard model. We analyzed the boundary behavior for various
helicity configurations in section 4.1 and wrote down the
corresponding BCFW recursion relations. In section 4.2, we present
explicit calculations to demonstrate our results.

There are two appendixes. In Appendix A we have present amplitudes
calculated directly by Feynman diagrams. Its role is to check
calculations did by BCFW recursion relation with boundary
contributions. In Appendix B, we have discussed the boundary
contributions for general $2l$ fermions in scalar QCD.

\section{The  $\la\phi ^{4}$ Theory}

In this section, we discuss our first example with nonzero boundary
contribution: the $\la\phi^4$ theory. We will analyze the boundary
behavior first, and then write down the BCFW recursion relation with
boundary contribution. As a comparison we have done same calculation
using the standard Feynman diagram method in Appendix A.

\subsection{The boundary behavior and BCFW relation}

Let us consider following BCFW deformation for massless $\la \phi^4$
theory
\begin{equation}
\lambda^{(i)}=\lambda^{(i)}+z \lambda^{(j)} \quad
\tilde{\lambda}^{(j)}=\tilde{\lambda}^{(j)}-z\tilde{\lambda}^{(i)}~~\label{eq:bcfwshift}
\end{equation}
Using the only nontrivial vertex $\la\phi^4$ to construct the
tree-level Feynman diagram, we found that all diagrams can be
divided into two categories: (A) particles $i,j$ are attached to
same vertex; (B) particles $i,j$ are attached to different vertexes.
For diagrams in category (B), there is at least one propagator on
the line connecting $i,j$ depending on $z$ linearly, i.e., we will
have factor ${1\over P^2- z\Spab{j|P|i}}$ in the expression. Thus
under the limit $z\to \infty$, expressions in category (B) will go
to zero, so they do not give boundary contributions.

Opposite to the category (B), since $i,j$ are attached to same
vertex, the whole expressions in category (A) do not depend on $z$
at all. In another word, there are nonzero boundary contributions
from category (A). By this simple analysis, we know that the
boundary contribution can be  calculated  by attaching the
lower-point tree level amplitudes to this vertex.

Having above analysis, we can immediately write down the BCFW
on-shell recursion relation for this simple theory as
\bea A= A_b+ A_{pole}~~~~\label{Phi4-BCFW-1}\eea
where $A_b$ as boundary contribution given by
\bea A_b & = & (-i\lambda)\sum_{ {\cal I^{\prime }\bigcup J^{\prime
}}=\{n\}\backslash \{i,j\}}A_{\cal I^{\prime }}\left(\{K_{\cal
I^{\prime }})\right)\frac{1}{P_{\cal I^{\prime }}^2}\frac{1}{P_{\cal
J^{\prime }}^2}A_{\cal J^{\prime }}\left(\{K_{\cal J^{\prime
}}\}\right)~~~~\label{Phi4-BCFW-3}\eea
and $A_{pole}$ as contributions from poles  given by standard
BCFW-form
\bea A_{pole} & = & \sum_{i\in{\cal I},j\not\in {\cal I}} A_{\cal
I}\left( \{K_{\cal I^{\prime }}\}, p_{i}(z_{\cal I}), -P_{\cal
I}(z_{\cal I}) \right)\frac{1}{P_{\cal I}^2}A_{\cal J}\left(
\{K_{\cal J^{\prime }}\}, p_{j}(z_{\cal I}), P_{\cal I}(z_{\cal I})
\right)~~~~\label{Phi4-BCFW-2}\eea
The expression (\ref{Phi4-BCFW-3})   just states the fact that set
${\cal I^{\prime }}$, set ${\cal J^{\prime }}$ and particles $i,j$
are attached to same vertex with coupling constant $-i\la$. There
two contributions can be represented by Figure\ref{Fig:division}
$(a)$ and $(b)$, where we have set $i,j=1,2$:

   \begin{figure}[hbt]
  \centering
  \includegraphics[viewport=150 595 492 713,clip]{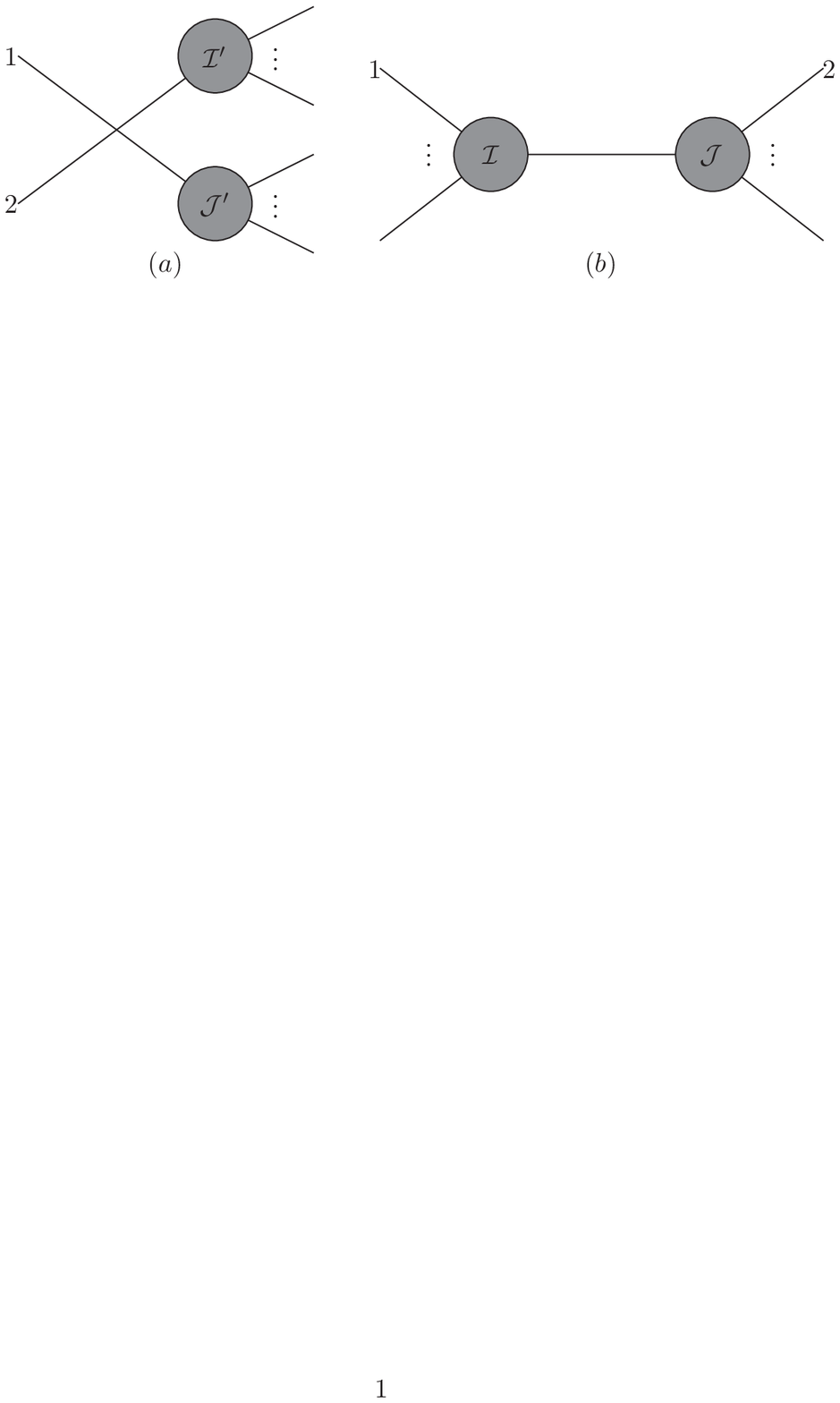}
  \caption{$(a)$ The contribution from boundary.
      $(b)$ The contribution from pole part. \label{Fig:division}}
\end{figure}

In following subsection, we will check the formula above using explicit
calculations. For simplicity, we will focus on the color ordered
case. For ordered case, something new is happening: when particles
$i,j$ have distance more than two, they will never be attached to
same vertex and the boundary contribution will be zero. Thus we can
check our result using $\Spab{1|2}$-shifting with boundary
contribution against the $\Spab{1|4}$-shifting without boundary
contribution. We want to emphasize that although for ordered case we
can have deformation  without boundary contribution, in real
calculation, we need to sum up all orderings, so formula with
boundary contribution will be unavoidable.

Having above explanation, in following calculations, we will write
down results from shifting $\Spab{1|2}$ and compare them with the
one from shifting $\Spab{1|4}$ as well as the one with direct
Feynman diagrams in Appendix A.

\subsection{The $\Spab{1|2}$ shifting}

Since for $\la\phi^4$ theory we have only quadruple vertex,
tree-level amplitudes with odd number of scalars are automatically
zero, thus we will consider six, eight and ten point amplitudes
only.

In this part we will use the $\Spab{1|2}$ shifting  given by
\bea \la_1(z)=\la_1+z \la_2,~~~~~\W\la_2(z)=\W\la_2-z\W\la_1\eea
thus we have following results.\\

{\bf Six-point amplitudes:}\\

First we consider the $\Spab{1|2}$ shifting.  It is easy to see that
there is only one figure contributing to pole part
\bea A_{6,pole}^{\Spab{1|2}}(1,\dots,6)&=&
A_4(5,6,\widehat{1},-\widehat{P})\frac{1}{P_{561}^2}A_4(\widehat{P},\widehat{2},3,4)
   =(-i\lambda)^2\left(\frac{1}{P_{156}^2}\right)
\eea
For the boundary part, there are two contributions
\bea
 A_{6,b}^{\Spab{1|2}} (1,\dots,6) & = &
 A_4(1,2,-P_1,-P_2)\left(\frac{1}{p_3^2}A_2(P_1,3)\right)
 \left(\frac{1}{P_{123}^2}A_4(P_2,4,5,6)\right)
  \nonumber\\
 && +A_4(1,2,-P_1,-P_2)\left(\frac{1}{P_{3,4,5}^2}A_4(P_1,3,4,5)\right)
   \left(\frac{1}{p_6^2}A_2(P_2,6)\right)
    \nonumber\\
   & = & (-i\lambda)^2 \left(\frac{1}{P_{123}^2}+\frac{1}{P_{126}^2}\right)
      \eea
where for simplicity we have defined the notation
$A_2(a,b)=\delta^4(p_a-p_b) p_a^2$, $P_{ijk}=p_i+p_j+p_k$. Putting
together we have
\bea
 A_6^{FD}(1,\dots,6)=(-i\lambda)^2\left(\frac{1}{P_{123}^2}
    +\frac{1}{P_{126}^2}+\frac{1}{P_{156}^2}\right)
\eea
  which agrees with the one from three Feynman diagrams.

For the shifting $\Spab{1|4}$, there is no boundary part but there
are three terms in pole part: $(123|456)$, $(612|345)$ and
$(561|234)$. Adding three terms together we get the same answer.\\

{\bf Eight-point amplitudes:}\\

There are two types of trees with three vertexes contributing at
this level: (A) type $(123|45|678)$ plus $Z_8$ cyclic ordering and
(B) type $(123|48|567)$ where $4,8$ at at the two sides of
propagators, plus $Z_4$ cyclic ordering. Adding them together we
have
\bea
 A_8^{FD}(1,\dots,8)&=&A_8^{FD(a)}+A_8^{FD(b)}
      \nonumber \\
         &=&(-i\lambda)^3\sum_{\sigma\in\mathbb{Z}_8}\left(\frac{1}{P_{\sigma(1)\sigma(2)\sigma(3)}^2
             P_{\sigma(6)\sigma(7)\sigma(8)}^2}+\frac{1}{2P_{\sigma(1)\sigma(2)\sigma(3)}^2P_{
                  \sigma(5)\sigma(6)\sigma(7)}^2}\right)
      \eea
where $FD$ means the result from direct Feynman diagrams.

Now we use the $\Spab{1|2}$-shifting. The pole contribution is given
by sum of following two terms (notice that the tree amplitude is
zero with odd number of external lines) :
\bea A_{8,pole}^{\Spab{1|2}}(1,2,\dots,8)
        &=&\widehat{A}_4(7,8,\widehat{1},-\widehat{P})\frac{1}{P^2_{178}}\widehat{A}_6(\widehat{P},\widehat{2},3,4,5,6)
                  +\widehat{A}_6(5,6,7,8,\widehat{1},-\widehat{P})\frac{1}{P^2_{234}}
           \widehat{A}_4(\widehat{P},\widehat{2},3,4)
                     \nonumber \\
        &=&{(-i\lambda)^3}\left[\frac{1}{P^2_{178}}\left(\frac{1}{\widehat{P}^2_{\widehat{2}34}}
                        +\frac{1}{{P}^2_{345}}+\frac{1}{{P}^2_{456}}\right)+
                        \left(\frac{1}{{P}^2_{567}}+\frac{1}{{P}^2_{678}}+
                        \frac{1}{\widehat{P}^2_{\widehat{1}78}}\right)
                        \frac{1}{P^2_{234}}\right]
      \eea
Using the locations of the poles
$z_1=-\frac{P^2_{178}}{\Spab{1|P_{178}|2}}$,
       $z_2=-\frac{P^2_{234}}{\Spab{1|P_{234}|2}}$, we can simplify
 \bea
  \frac{1}{P^2_{178}\widehat{P}^2_{\widehat{2}34}}+\frac{1}{\widehat{P}^2_{\widehat{1}78}P^2_{234}}
         &=&\frac{1}{P^2_{178}P^2_{234}}\left(\frac{1}{1-\frac{z_1}{z_2}}+\frac{1}{1-\frac{z_2}{z_1}}\right)
             =\frac{1}{P^2_{178}P^2_{234}}~,
      \eea
where we have used  the identity
$\frac{1}{1-\frac{z_1}{z_2}}+\frac{1}{1-\frac{z_2}{z_1}}=1$. In
fact, there is a general identity
      \bea\label{eq:id1}
        \sum_{i=1}^n\ \prod_{\substack{j=1 \\ j \ne i}}^n \frac{1}{1-\frac{z_i}{z_j}}=1,
      \eea
which will be  useful also in  our ten-point calculation.

For  boundary part, there are following three splitting $(3|45678)$,
$(345|678)$ and $(34567|8)$. Adding them up we have
\bea
 A_{8,b}^{\Spab{1|2}} (1,2,\dots,8)
        =&{(-i\lambda)}^3\Bigg[\frac{1}{P^2_{123}}
        \left(\frac{1}{P^2_{456}}+\frac{1}{P^2_{567}}+\frac{1}{P^2_{678}}\right)
    +\frac{1}{P^2_{812}}\left(\frac{1}{P^2_{345}}+\frac{1}{P^2_{456}}+\frac{1}{P^2_{567}}\right)
                    +\frac{1}{P^2_{345}P^2_{678}}
                    \Biggr]
\eea
It is easy to check that add the pole part and boundary part we
indeed reproduce the result from Feynman diagrams.

Now we move to the shifting $\Spab{1|4}$. There is no boundary part,
but for the pole part, there are following six terms: $(123|45678)$,
$(812|34567)$, $(781|23456)$, $(78123|456)$, $(67812|345)$ and
$(56781|234)$. Adding them up, we get again the same answer.

It should be interesting to compare terms we have added up in each
method. For Feynman diagram method, there are $8+4=12$ terms. For
$\Spab{1|2}$-shifting there are $2+3=5$ terms while for
$\Spab{1|4}$-shifting there are $6$ terms. Different method has
given different combinations of various propagators. \\

{\bf Ten-point amplitudes:}\\

For this case, there are several topologies for tree amplitudes as
shown in Appendix A. The result can be summarized with  cyclic
ordering as (\ref{10-point}). There are four kinds of diagrams with
$Z_{10}$ cyclic ordering and another three, $Z_{5}$ cyclic ordering,
so there are total $55$ terms.

For the $\Spab{1|2}$-shifting,  the boundary part has following four
terms: $(3|456789(10))$, $(345|6789(10))$, $(34567|89(10))$ and
$(3456789|(10))$. The result is given by
 \bea
 && \    A_{10,b}^{\Spab{1|2}}(1,2,...,10)\nonumber \\
  &=&A_4(1,2,-P_1,-P_2)\left(\frac{1}{p^2_3}A_2(P_1,3)\right)\left(\frac{1}{P_{123}^2}
           A_8(P_2,4,...,10)\right)\nonumber \\
        && +A_4(1,2,-P_1,-P_2)\left(\frac{1}{P^2_{345}}A_4(P_1,3,4,5)\right)\left(\frac{1}{P^2_{12345}}
           A_6(P_2,6,...,10)\right)\nonumber \\
        && +A_4(1,2,-P_1,-P_2)\left(\frac{1}{p^2_{34567}}A_6(P_1,3,...,7)\right)\left(\frac{1}{P_{89(10)}^2}
           A_4(P_2,8,9,10)\right)\nonumber \\
        && +A_4(1,2,-P_1,-P_2)\left(\frac{1}{p^2_{(10)12}}
           A_8(P_1,3,...,9)\right)\left(\frac{1}{p_{10}^2}A_2(P_2,10)\right)\nonumber \\
        &=&(-i\lambda)^4\frac{1}{P_{123}^2}\left(\sum_{\sigma\in\mathbb{Z}_8}
           \left(\frac{1}{P^2_{\sigma(3)\sigma(4)\sigma(5)}P^2_{\sigma(6)\sigma(7)\sigma(8)}}
          +\frac{1}{2P^2_{\sigma(3)\sigma(4)\sigma(5)}P^2_{\sigma(7)\sigma(8)\sigma(9)}}\right)\right)\nonumber \\
        && +\frac{1}{P^2_{345}P^2_{12345}}\left(\frac{1}{P^2_{678}}+\frac{1}{P^2_{789}}+\frac{1}{P^2_{89(10)}}\right)+
           \frac{1}{P^2_{34567}P^2_{89(10)}}\left(\frac{1}{P^2_{345}}+\frac{1}{P^2_{456}}+\frac{1}{P^2_{567}}\right)\nonumber\\
        && +\frac{1}{P_{12(10)}^2}\sum_{\sigma\in\mathbb{Z}_8}
           \left(\frac{1}{P^2_{\sigma(3)\sigma(4)\sigma(5)}P^2_{\sigma(6)\sigma(7)\sigma(8)}}
           +\frac{1}{P^2_{\sigma(3)\sigma(4)\sigma(5)}P^2_{\sigma(7)\sigma(8)\sigma(9)}}\right)\nonumber \\
   \eea
The  pole part is given by the sum of three terms
 \bea
 && A_{10,pole}^{\Spab{1|2}} (1,2,...10)\nonumber \\
   &=&\widehat{A_4}(9,10,\widehat{1},-P)\frac{1}{P^2_{9(10)1}}\widehat{A_8}(P,\widehat{2},3,4,5,6,7,8)
   \nonumber\\
      && +\widehat{A_6}(7,8,9,10,\widehat{1},-P)\frac{1}{P^2_{2345}}\widehat{A_6(}P,\widehat{2},3,4,5,6)
         +\widehat{A_8}(5,6,7,8,9,10,\widehat{1},-P)\frac{1}{P^2_{234}}\widehat{A_4}(P,\widehat{2},3,4)\nonumber \\
      &=&(-i\lambda)^4\frac{1}{P_{9(10)1}^2}\left(\sum_{\sigma\in\mathbb{Z}_8}
          \left(\frac{1}{\widehat{P}^2_{\sigma(2)\sigma(3)\sigma(4)}\widehat{P}^2_{\sigma(5)\sigma(6)\sigma(7)}}+
          \frac{1}{2\widehat{P}^2_{\sigma(2)\sigma(3)\sigma(4)}
          \widehat{P}^2_{\sigma(6)\sigma(7)\sigma(8)}}\right)\right)\nonumber\\
      && +(\frac{1}{\widehat{P}^2_{789}}+\frac{1}{\widehat{P}^2_{89(10)}}
         +\frac{1}{\widehat{P}^2_{9(10)1}})\frac{1}{P^2_{23456}}(\frac{1}{\widehat{P}^2_{234}}
         +\frac{1}{\widehat{P}^2_{345}}+\frac{1}{\widehat{P}^2_{456}})\nonumber \\
      && +\frac{1}{P_{234}^2}\left(\sum_{\sigma\in\mathbb{Z}_8}
         \left(\frac{1}{\widehat{P}^2_{\sigma(5)\sigma(6)\sigma(7)}
         \widehat{P}^2_{\sigma(8)\sigma(9)\sigma(10)}}
         +\frac{1}{2\widehat{P}^2_{\sigma(5)\sigma(6)\sigma(7)}
         \widehat{P}^2_{\sigma(9)\sigma(10)\sigma(1)}}\right)\right)
         \nonumber \\
      \eea
Using (\ref{eq:id1}) we can show following identity:
      \bea\label{eq:red12}
      &&(-i\la)^4 [\frac{1}{P^2_{9(10)1}}
        (\frac{1}{\widehat{P}^2_{\widehat{2}34}}\frac{1}{P^2_{567}}
        +\frac{1}{P^2_{456}}\frac{1}{\widehat{P}^2_{\widehat{2}3456}}
        +\frac{1}{\widehat{P}^2_{\widehat{2}3456}}\frac{1}{\widehat{P}^2_{\widehat{2}34}}
        +\frac{1}{\widehat{P}^2_{\widehat{2}34}}\frac{1}{P^2_{567}}
        +\frac{1}{P^2_{345}}\frac{1}{\widehat{P}^2_{\widehat{2}3456}})\nonumber \\
      &&+\frac{1}{P^2_{234}}(\frac{1}{P^2_{678}}\frac{1}{\widehat{P}^2_{9(10)\widehat{1}}}
        +\frac{1}{\widehat{P}^2_{9(10)\widehat{1}}}\frac{1}{\widehat{P}^2_{\widehat{2}3456}}
        +\frac{1}{\widehat{P}^2_{\widehat{2}3456}}\frac{1}{P^2_{789}}
        +\frac{1}{P^2_{567}}\frac{1}{\widehat{P}^2_{9(10)\widehat{1}}}
        +\frac{1}{P^2_{89(10)}}\frac{1}{\widehat{P}^2_{\widehat{2}3456}})\nonumber\\
      &&+\frac{1}{\WH{P}^2_{9(10)\WH{1}}}\frac{1}{P^2_{23456}}(\frac{1}{\WH{P}^2_{\WH{2}34}})]
\nn
    & = &   (-i\la)^4 [\frac{1}{P^2_{9(10)1}}
        (\frac{1}{P^2_{234}}\frac{1}{P^2_{567}}
        +\frac{1}{P^2_{456}}\frac{1}{P^2_{23456}}
        +\frac{1}{P^2_{23456}}\frac{1}{P^2_{234}}
        +\frac{1}{P^2_{234}}\frac{1}{P^2_{567}}
        +\frac{1}{P^2_{345}}\frac{1}{P^2_{23456}})
      \eea
and then we can show that the result is same as given by
$A^{FD}_{10}$.

For shifting $\Spab{1|4}$, there are nine terms from the recursion
relations. Summing it up with some algebra we see that it reproduce
the right answer.

Again, we count terms from different methods. The Feynman diagrams
give $55$ terms. The $\Spab{1|2}$ shifting gives $4+3=7$ terms while
the $\Spab{1|4}$ shifting gives $9$ terms.

\section{Boundary BCFW and auxiliary field}

As we have discussed, for $\la\phi^4$ theory, the boundary
contribution is not zero for BCFW deformation (here we means the
general unordered case). However, as presented in \cite{Paolo:2007},
by  introducing  a massive  auxiliary field $\chi$, it is possible
to rewrite the $\la\phi^4$ theory into another form where
three-particle  BCFW deformation without boundary contribution does
exists. In this section, we will explore the relation between
auxiliary field method and the boundary BCFW method for $\la\phi^4$
theory.

The new  Lagrangian with auxiliary field is given by
\begin{equation}
\mathcal{L}(\phi,\chi)\:=
  \frac{1}{2}\left(\partial_{\mu}\phi\right)\left(\partial^{\mu}\phi\right)+
  \frac{1}{2}\left(\partial_{\mu}\chi\right)\left(\partial^{\mu}\chi\right)
  -\frac{1}{2}m_{\chi}^{2}\chi^{2}
  -g\chi\phi^{2}.
\end{equation}
The theory is not the $\la\phi^4$ theory, but under some limit, we
can recover late. The limit is the large mass limit, where
 $\chi$ is not excited, so we can use the equation of motion (where
the kinematic part has been set to zero) $m_\chi^2 \chi+g \phi^2=0$
to solve $\chi$ and then  put it back  to the Lagrangian to get
\begin{equation}
\mathcal{L}(\phi)\:=
  \frac{1}{2}\left(\partial_{\mu}\phi\right)\left(\partial^{\mu}\phi\right)-{\la\over
  4!}
  \phi^{4}.
\end{equation}
where to match up the coupling constant, we need to set ${g^2\over 2
m_\chi^2}={\la\over 4!}$.

The Lagrangian ${\cal L}(\phi,\chi)$ is on-shell constructible
 without boundary contribution under following three-particle
BCFW deformation
\bea \tilde{\lambda}^{(1)}(z) & = & \tilde{\lambda}^{(1)}-
 z\left(\frac{[1,3]}{[2,3]}\tilde{\lambda}^{(2)}
       +\frac{[1,3]}{[3,4]}\tilde{\lambda}^{(4)}\right)\nn
\lambda^{(2)}(z)& = &
\lambda^{(2)}+z\frac{[1,3]}{[2,3]}\lambda^{(1)},~~~~
\lambda^{(4)}(z)=\lambda^{(4)}+z\frac{[1,3]}{[3,4]}\lambda^{(1)}.
~~~~\label{aushift}\eea
but ${\cal L}(\phi)$ has boundary contribution under same
deformation. In another word, for ${\cal L}(\phi,\chi)$ theory the
BCFW recursion relation for $n$ $\phi$ scalars is given by
\bea \W A_n & = & \sum_{i\in I,~ 2~or~4\in J} \W A_{I}( I(z),
P_\phi){1\over P^2}\W A_J(J(z),-P_\phi)\nn & & +\sum_{i\in I,~
2~or~4\in J} \W A_{I}( I(z), P_\chi){1\over P^2-m^2_\chi} \W
A_J(J(z),-P_\chi)~,~~~~~~\label{A-Phi-Chi}\eea
where the first term has $\vev{\phi\phi}$ propagator in middle and
the second term, has $\vev{\chi\chi}$ propagator. For  ${\cal
L}(\phi)$ theory the corresponding  recursion relation is modified
to
\bea A_n & = & \sum_{i\in I,~ 2~or~4\in J} A_{I}( I(z),
P_\phi){1\over P^2} A_J(J(z),-P_\phi)+
A_{n,b}~~~~~~\label{A-Phi}\eea
where $A_b$ is the contribution from boundary. Comparing these two
formula (\ref{A-Phi-Chi}) and (\ref{A-Phi}), we find that first term
of both formula is, in fact, identical. Thus $\W A_n=A_n$ is
equivalent to the condition that the second term of $\W A_n$, which
is  provided by the auxiliary propagator, is equal to the boundary
part of $A_n$. Now we show this is true in remaining part of this
section.\\

{\bf The boundary part of  (\ref{A-Phi}):}\\

\begin{figure}[hbt]
  \centering
  \includegraphics[viewport=122 548 592 705,clip]{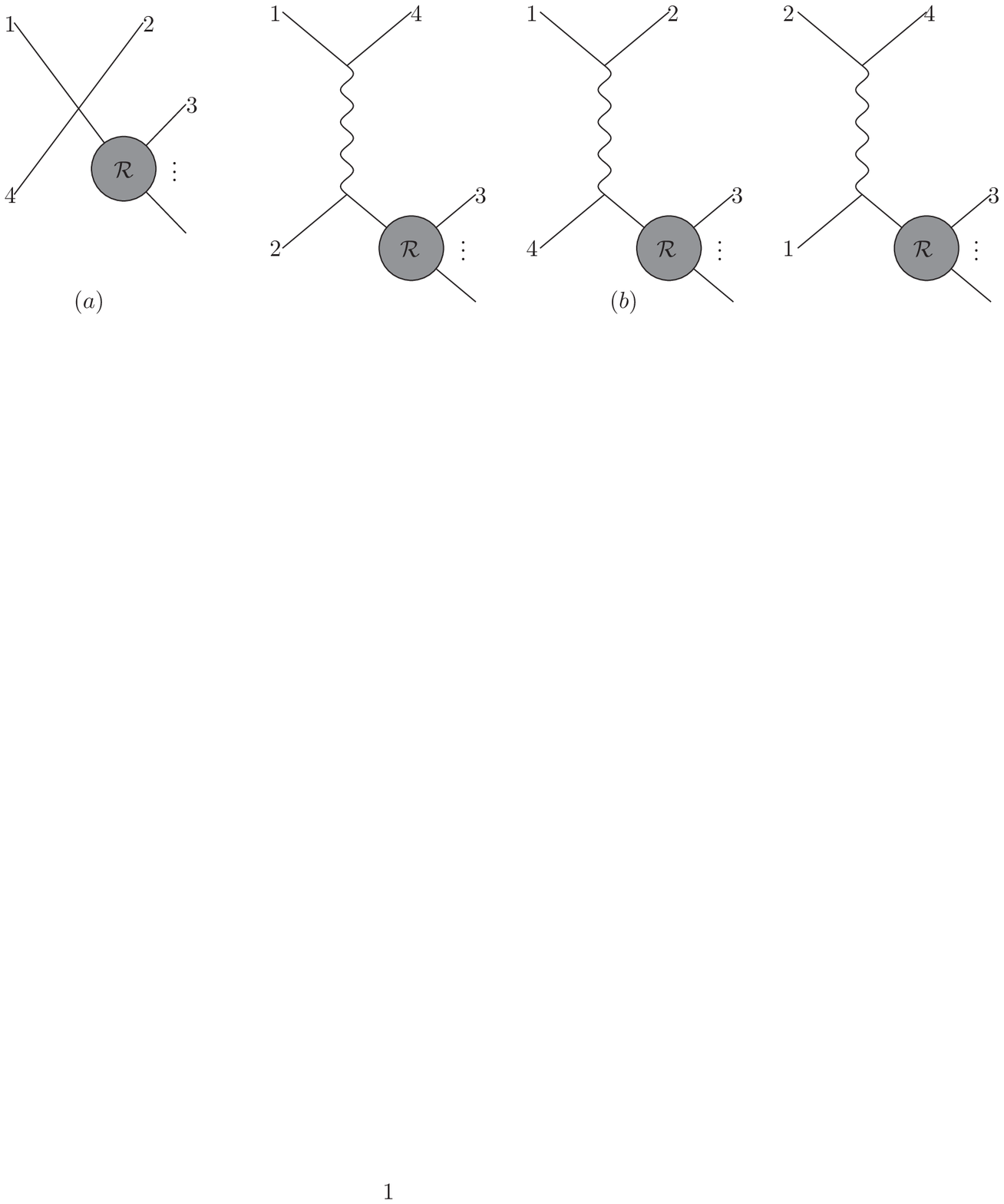}
  \caption{pure scalar case: $(a)$ boundary term figure
  $(b)$ boundary term figure in view of auxiliary field\label{Fig:scalar}}
\end{figure}

Just by checking the Feynman diagrams, it is easy to identify which
kind of Feynman diagrams contributes the boundary term. It is
nothing, but the one where particles $1,2,4$ are attached to same
vertex, as shown in Figure\ref{Fig:scalar} $(a)$.

{\bf The second term of (\ref{A-Phi-Chi}):}\\

For this part, we need to use the amplitudes with one $\chi$ field,
so it is important to know the $g, m_\chi$ power dependence of these
amplitudes. First it is easy to know  that the amplitude of $m$
$\phi$ scalars and one $\chi$ scalar is zero when $m=odd$ while when
$m=even$, it is not zero. Assuming there are $V$ triple-vertex,
$I_1$ $\vev{\phi\phi}$ propagators and $I_2$ $\vev{\chi\chi}$
propagators, by some simple arguments we have
\bea I_1=I_2={m\over 2}-1,~~~~V=m-1~, \eea
thus the large mass limit is given by
\bea A(m \phi,\chi)\sim {g^V\over (m_\chi^2)^{m/2-1}}\sim
\la^{m-1\over 2}m_\chi~.\eea

Now we consider the amplitude under $z$-deformation where $z$ is
solved by  $P^2(z)-m_\chi^2=0$, i.e., $z_\a\sim m_\chi^2$ for large
mass limit. With the appearance of $z$, propagators will have
different large mass behavior than the one without $z$-deformation.
 For $\vev{\chi\chi}$ propagator
since $P^2(z_\a)-m_\chi^2\sim m_\chi^2$,  the large mass behavior is
same before and after $z$-deformation. Opposite to that, the
$\vev{\phi\phi}$ propagator will be $P^2(z)\sim m_\chi^2$ after
$z$-deformation and $P^2\to (m_\chi^2)^0$ before the
$z$-deformation. Putting this back, we have
\bea A(m \phi,\chi,z_\a)\sim \la^{m-1\over 2}m_\chi^{1-2
t}~,~~~\label{Am-z}\eea
where $t$ is the number of $\vev{\phi\phi}$ propagators affected by
$z$-deformation. Applying (\ref{Am-z}) to the second term of
(\ref{A-Phi-Chi}) we have
\bea \la^{m_L-1\over 2}m_\chi^{1-2 t_L} {1\over m^2_\chi}
\la^{m_R-1\over 2}m_\chi^{1-2 t_R} \sim \la^{n-2\over 2}
(m^2_\chi)^{-2(t_L+t_R)}~,\eea
 which is not zero under large
mass limit when and only when $t_L=t_R=0$.

What are these nonzero contributions with $t_L=t_R=0$ under our
triple deformation? They are nothing, but the one given by Figure
\ref{Fig:scalar} $(b)$. It is also easy to see that they correspond
exactly to the boundary part of (\ref{A-Phi}).

Thus we have shown that how the boundary contribution can be
transferred into contribution from auxiliary fields under triple
deformation.

\section{The scalar QCD theory}

In this section, we consider another example where scalar and
fermion interacting by Yukawa coupling form $\O \psi \phi \psi$.
Similar interaction terms are presented in Standard Model, so our
example will have potential applications for practical calculations.

As in previous section, we will discuss the boundary behavior first
and then write down the BCFW recursion relation with boundary
contributions. After that we  do several concrete calculations to
demonstrate our method. For simplicity, our attention will focus on
color ordered amplitudes with
 exactly two fermions and $n$ scalars.  Other situations can be
 discussed similarly.

\subsection{The analysis of Feynman diagrams}

\begin{figure}[hbt]
  \centering
  \includegraphics[viewport=160 607 465 705,clip]{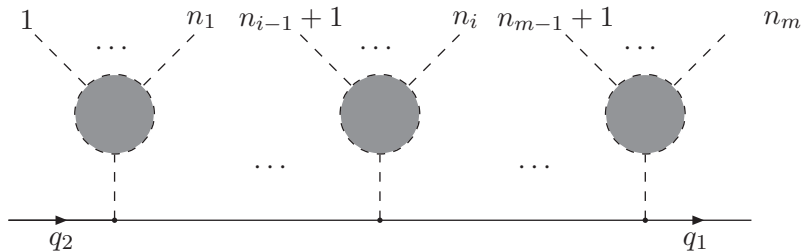}
  \caption{General Feynman diagrams. \label{Fig:general}}
\end{figure}

For ordered amplitudes with two fermions and $n$ scalars, we use
$q_1,q_2$ to denote momenta of fermions and $p_1,...,p_n$, the
momenta of scalars, thus the ordered amplitude is denoted by
$A(q_1,p_1,...,p_n,q_2)$. By inspecting the general Feynman diagram
given in Figure \ref{Fig:general}, we see that there is one common
feature: {\sl  a single  fermionic line connecting these two
fermions while other scalars are attached through Yukawa coupling at
same side}. Using the fermionic propagator ${i\not{p}\over p^2}$,
the amplitude can be written as
\bea  A= \sum_{diagrams}{\cal S}_i {\cal Q}_i  ~~~~\label{A-SQ}\eea
where $S_i$ is contribution from scalar part and $Q_i$ is the form
\bea {\cal Q}(q_1^-, q_2^+; R_1,...,R_m)\sim
i^m{\Spab{1|R_1|R_2|...|R_m|2}\over R_1^2 R_2^2 .... R_m^2} \eea
where we have assumed the helicity of $q_1,q_2$ is $(-,+)$ and there
are $m$ fermionic propagators along the line. In fact, it is easy to
see that when $h_{q_1}=h_{q_2}$, to get nonzero amplitudes we must
have even number of fermionic propagators (i.e., $m=even$) while
when $h_{q_1}=-h_{q_2}$, to get nonzero amplitudes we must have odd
number of fermionic propagators (i.e., $m=odd$).

Now we  introduce following $\Spab{1|2}$-deformation  between the
two fermions
\bea\lambda^{(1)}=\lambda^{(1)}+z \lambda^{(2)}
\quad\tilde{\lambda}^{(2)}=\tilde{\lambda}^{(2)}-z\tilde{\lambda}^{(1)}
\eea
It is easy to see from Fig \ref{Fig:general} that all ${\cal S}_i$
factor in (\ref{A-SQ}) do not depend on $z$ and the only
$z$-dependence  is inside ${\cal Q}_i$. With different helicity
configurations, the discussion will be a little different, so we
consider it case by case.\\

{\bf The helicity $(h_{q_1}, h_{q_2})=(+,+)$:}\\

In this case the number of propagator should be even and we have
following two cases: (A) $m=0$; (B) $m\geq 2$. For case (B), we have
\bea & & {\cal Q}(q_1^+, q_2^+; R_1,...,R_m)\sim
i^m{\Spbb{1|(q_1+R_1+z \la_2\W\la_1)(\prod_{j=1}^{m-2} (q_1+R_j+z
\la_2\W\la_1))(q_1+R_m+z \la_2\W\la_1)|\W\la_2-z \W\la_1}\over
(q_1+R_1+z \la_2\W\la_1))^2(\prod_{j=1}^{m-2} (q_1+R_j+z
\la_2\W\la_1)^2)(q_1+R_m+z \la_2\W\la_1)^2 }\nn
 & = & i^m {\Spbb{1|(q_1+R_1)(\prod_{j=1}^{m-2} (q_1+R_j+z \la_2\W\la_1))(q_1+R_m+z
\la_2\W\la_1)|\W\la_2-z \W\la_1}\over (q_1+R_1+z
\la_2\W\la_1))^2(\prod_{j=1}^{m-2} (q_1+R_j+z
\la_2\W\la_1)^2)(q_1+R_m+z \la_2\W\la_1)^2 }\nn
 & = & i^m {\Spbb{1|(q_1+R_1)(\prod_{j=1}^{m-2} (q_1+R_j+z \la_2\W\la_1))
 (q_1+R_m)|\W\la_2-z \W\la_1}\over (q_1+R_1+z
\la_2\W\la_1))^2(\prod_{j=1}^{m-2} (q_1+R_j+z
\la_2\W\la_1)^2)(q_1+R_m+z \la_2\W\la_1)^2 }\nn
 & &+ i^m {\Spbb{1|(q_1+R_1)(\prod_{j=1}^{m-2} (q_1+R_j+z \la_2\W\la_1))(z
\la_2\W\la_1)|\W\la_2}\over (q_1+R_1+z
\la_2\W\la_1))^2(\prod_{j=1}^{m-2} (q_1+R_j+z
\la_2\W\la_1)^2)(q_1+R_m+z \la_2\W\la_1)^2 } \eea
which goes zero under the $z\to \infty$ limit since each term has
$(m-1)$ $z$ in numerator and $m$ $z$ in denominator. For the case
(A) we have
\bea {\cal Q}(q_1^+, q_2^+)=\Spbb{1|2-z 1}=\Spbb{1|2}\eea
which is independent of $z$.

From above analysis we see that under our $\Spab{1|2}$ shifting,
there is nonzero boundary contribution and it is purely given by
diagrams of case (A). Thus it is easy to write down the BCFW recursion relation
with boundary term as
\bea & & A_{n+2}(q_1^+;p_1,...,p_n;q_2^+) \nn
& = &
\sum_{i=1,h=\pm}^{n-1}
A_{i+2}(q_1^+(z_i);p_1,...,p_i;q_i^h(z_i)){1\over (q_1+\sum_{j=1}^i
p_j)^2}A_{n-i+2}(-q_i^{-h}(z_i);p_{i+1},...,p_n;q_2^+(z_i))\nn
& & + {(-i g)\Spbb{1|2}\over (\sum_{i=1}^n p_i)^2}
A_{n+1}(p_1,...,p_n,P_{\phi}) ~~~~\label{12++BCFW}\eea
where $A_{n+1}$ is  the amplitude of $(n+1)$ pure scalars.  The
expression can also be represented by following Figure
\ref{Fig:total}
\begin{figure}[htb]
  \centering
  \includegraphics[viewport=160 614 240 710,clip]{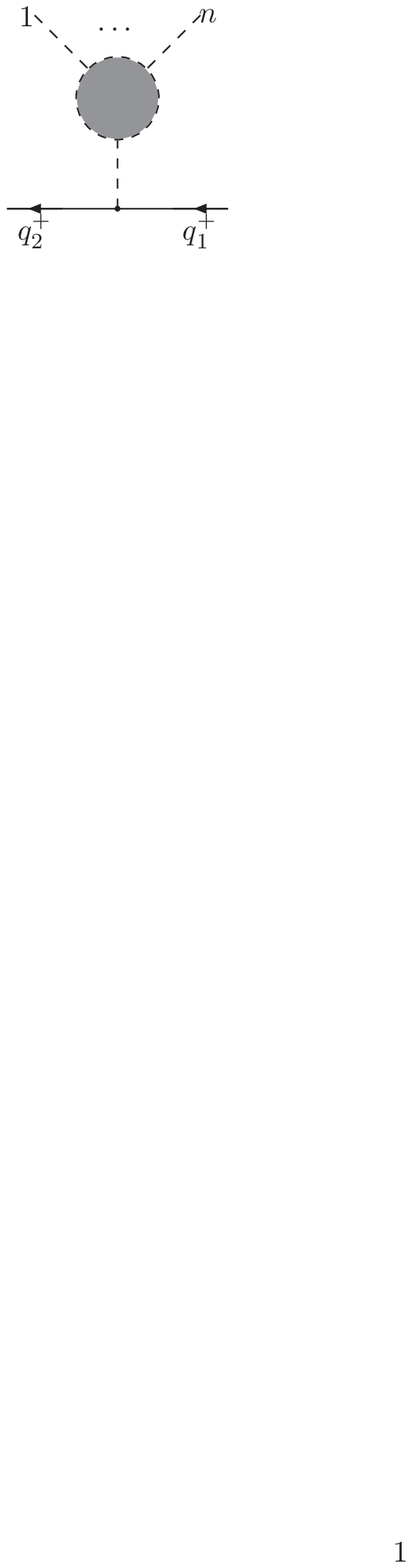}
  \includegraphics[viewport=160 605 465 705,clip]{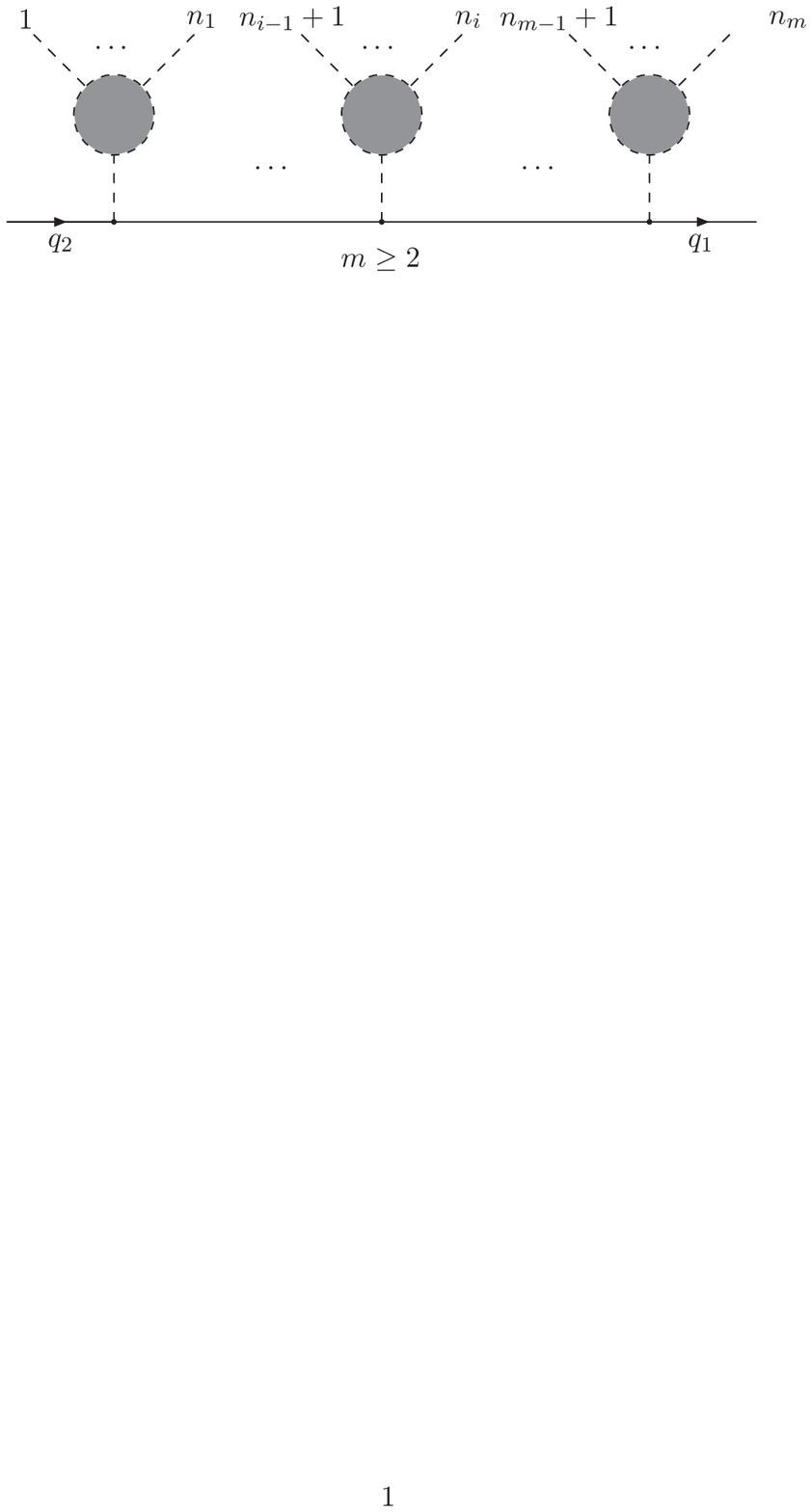}
  \caption{Amplitude in total\label{Fig:total}}
\end{figure}
\\

{\bf The helicity $(h_{q_1}, h_{q_2})=(-,-)$:}\\

In this case the number of propagator should again be even and we
have following two cases: (A) $m=0$; (B) $m\geq 2$. For case (B), we
have
\bea & & {\cal Q}(q_1^-, q_2^-; R_1,...,R_m)
\nn
&\sim&i^m{\Spaa{\la_1+z\la_2|(q_1+R_1+z \la_2\W\la_1)(\prod_{j=1}^{m-2}
(q_1+R_j+z \la_2\W\la_1))(q_1+R_m+z \la_2\W\la_1)|\la_2}\over
(q_1+R_1+z \la_2\W\la_1))^2(\prod_{j=1}^{m-2} (q_1+R_j+z
\la_2\W\la_1)^2)(q_1+R_m+z \la_2\W\la_1)^2 }\nn
 & = & i^m {\Spaa{\la_1+z\la_2|(q_1+R_1+z \la_2\W\la_1
 )(\prod_{j=1}^{m-2} (q_1+R_j+z \la_2\W\la_1))(q_1+R_m)|\la_2}\over (q_1+R_1+z
\la_2\W\la_1))^2(\prod_{j=1}^{m-2} (q_1+R_j+z
\la_2\W\la_1)^2)(q_1+R_m+z \la_2\W\la_1)^2 }\nn
& = & i^m {\Spaa{\la_1+z\la_2|(q_1+R_1
 )(\prod_{j=1}^{m-2} (q_1+R_j+z \la_2\W\la_1))(q_1+R_m)|\la_2}\over (q_1+R_1+z
\la_2\W\la_1))^2(\prod_{j=1}^{m-2} (q_1+R_j+z
\la_2\W\la_1)^2)(q_1+R_m+z \la_2\W\la_1)^2 }\nn
& + & i^m {\Spaa{\la_1|(z \la_2\W\la_1
 )(\prod_{j=1}^{m-2} (q_1+R_j+z \la_2\W\la_1))(q_1+R_m)|\la_2}\over (q_1+R_1+z
\la_2\W\la_1))^2(\prod_{j=1}^{m-2} (q_1+R_j+z
\la_2\W\la_1)^2)(q_1+R_m+z \la_2\W\la_1)^2 }\nn
\eea
which goes zero under the $z\to \infty$ limit since each term has
$(m-1)$ $z$ in numerator and $m$ $z$ in denominator.  For the case
(A) we have
\bea {\cal Q}(q_1^-, q_2^-)=\Spaa{1+z 1|2}=\Spaa{1|2}\eea
which is independent of $z$. Thus under our $\Spab{1|2}$ shifting,
there is nonzero boundary contribution and it is purely given by
diagrams of case (A). Thus it is easy to write down the BCFW
recursion relation with boundary term as
\bea & & A_{n+2}(q_1^-;p_1,...,p_n;q_2^-) \nn
& = & \sum_{i=1,h=\pm}^{n-1}
A_{i+2}(q_1^-(z_i);p_1,...,p_i;q_i^h(z_i)){1\over (q_1+\sum_{j=1}^i
p_j)^2}A_{n-i+2}(-q_i^{-h}(z_i);p_{i+1},...,p_n;q_2^-(z_i))\nn
& & + {(-i g)\Spaa{1|2}\over (\sum_{i=1}^n p_i)^2}
A_{n+1}(p_1,...,p_n,P_{\phi}) ~~~~\label{12--BCFW}\eea
where $A_{n+1}$ is  the amplitude of $(n+1)$ pure scalars. It is obvious that
$(+,+)$ is conjugated to $(-,-)$.\\

{\bf The helicity $(h_{q_1}, h_{q_2})=(+,-)$:}\\

In this case the number of propagator should  be odd, i.e.,   we
have at least one propagator. This case is, in fact, simpler than
previous two cases. The general expression of ${\cal Q}$ should be
\bea & & {\cal Q}(q_1^+, q_2^-; R_1,...,R_m)\sim
i^m{\Spba{\W\la_1|(q_1+R_1+z \la_2\W\la_1)(\prod_{j=1}^{m-1}
(q_1+R_j+z \la_2\W\la_1))|\la_2}\over (q_1+R_1+z
\la_2\W\la_1))^2(\prod_{j=1}^{m-2} (q_1+R_j+z
\la_2\W\la_1)^2)(q_1+R_m+z \la_2\W\la_1)^2 }\nn
 & = & i^m{\Spba{\W\la_1|(q_1+R_1)(\prod_{j=1}^{m-1}
(q_1+R_j+z \la_2\W\la_1))|\la_2}\over (q_1+R_1+z
\la_2\W\la_1))^2(\prod_{j=1}^{m-2} (q_1+R_j+z
\la_2\W\la_1)^2)(q_1+R_m+z \la_2\W\la_1)^2 }
\eea
which goes zero under the $z\to \infty$ limit since each term has
$(m-1)$ $z$ in numerator and $m$ $z$ in denominator.  In other word,
with this helicity configuration and the choice of BCFW deformation,
the boundary contribution is zero and we have familiar BCFW
recursion relation which is given by
\bea & & A_{n+2}(q_1^+;p_1,...,p_n;q_2^-) \nn
& = & \sum_{i=1,h=\pm}^{n-1}
A_{i+2}(q_1^+(z_i);p_1,...,p_i;q_i^h(z_i)){1\over (q_1+\sum_{j=1}^i
p_j)^2}A_{n-i+2}(-q_i^{-h}(z_i);p_{i+1},...,p_n;q_2^-(z_i))~.
 ~~~~\label{12+-BCFW}\eea
It is worth to emphasize that although $(h_1,h_2)=(+,-)$ case does
not have boundary contributions,  the sub-amplitudes in the
recursive calculation will meet the helicities configurations
$(+,+)$, $(-,-)$ and $(-,+)$, thus the boundary contributions have
been included implicitly through these sub-amplitudes. \\

{\bf The helicity $(h_{q_1}, h_{q_2})=(-,+)$:}\\

In this case the number of propagator should be odd and we should
have at least one propagator. However, unlike the previous case
where boundary contribution is zero, current one is the most
complicated one and we should divide diagrams into three cases: (A)
$m=1$; (B) $m= 3$; (C) $m\geq 5$. Let us first show that the case
(C) does not give boundary contributions. The observation we will
use  is that when $z\la_2\W\la_1$ are nearby, its contribution is
zero, i.e., $\Spaa{\a|z\la_2\W\la_1|z\la_2\W\la_1|\b}=0$. Using
this, when we have five $(q_1+R_i+z\la_2 \W\la_1)$ factors in a row,
we can expand it into the power of $z$ as (it is worth to remember
that there are five corresponding propagators with ${1\over z}$
dependence)
\bean & &  (q_1+R_1+z\la_2 \W\la_1)(q_1+R_2)(q_1+R_3+z\la_2
\W\la_1)(q_1+R_4)(q_1+R_5+z\la_2 \W\la_1)\nn
& & +(q_1+R_1)(q_1+R_2+z\la_2 \W\la_1)(q_1+R_3)(q_1+R_4+z\la_2
\W\la_1)(q_1+R_5) +{\cal O}(z)\eean
For the first term, when we contract with spinor as
$\Spab{\la_1+z\la_2|(q_1+R_1+z\la_2 \W\la_1)|\b}$, we have
$\Spab{\la_1+z\la_2|(q_1+R_1)|\b}+\Spab{\la_1|(z\la_2 \W\la_1)|\b}$,
thus all terms are at most ${1\over z}$ order.

Having established that the case (C) does not give boundary
contribution, we move to case (A) and (B). For case (A) the general
result should be
\bea I_A & = & (-ig)^2 \sum_{i=1}^{n-1} {\Spab{\la_1+z\la_2| (q_1+
R_i+z \la_2\W\la_1)|\W\la_2-z\W\la_1}\over (q_1+ R_i+z
\la_2\W\la_1)^2} {A_{i,\phi}\over R_i^2
(q_1+q_2+R_i)^2}~,~~~\label{IA-exp}\eea
where $R_i=p_1+...+p_i$ and $A_{i,\phi}$ is the contribution from
scalars as
\bea  A_{i,\phi} & = & A_{i+1}(p_1,...,p_i, p_{R_i})
A_{n-i+1}(p_{i+1},...,p_n,p_{R_2} )\eea
Taking the residue of $I_A$ around $z=\infty$ we will get
\bea B[I_A] & = & (-ig)^2\sum_{i=1}^{n-1} { (R_i^2+ 2R_i\cdot
q_2)\over \Spab{2|R|1}} {A_{i,\phi}\over R_i^2
(q_1+q_2+R_i)^2}~.~~~~\label{B-IA}\eea

For the case (B), first we consider which term gives nonzero
contributions. Expanding the product of three
$(q_1+R_i+z\la_2\W\la_1)$ we find that following five terms:
\bean & & \Spab{\a|(q_1+R_1)(q_2+R_2)(q_3+R_3)|\b}+
\Spab{\a|(z\la_2\W\la_1)(q_2+R_2)(q_3+R_3)|\b}\nn & & +
\Spab{\a|(q_1+R_1)(q_2+R_2)(z\la_2\W\la_1)|\b}\nn
& & + \Spab{\a|(q_1+R_1)(z\la_2\W\la_1)(q_3+R_3)|\b}+
\Spab{\a|(z\la_2\W\la_1)(q_2+R_2)(z\la_2\W\la_1)|\b}\eean
with $\a=\la_1+z\la_2,\b=\W\la_2-z\W\la_1$. Remembering the ${1\over
z^3}$ factor from three propagators, we see that except the fourth
term, all other terms will vanish under the limit $z\to \infty$. In
another word, only the fourth term gives nonzero contribution.

Having identified the term, we write down its expression as
\bea I_B & = & (-ig)^4 \sum_{i_1+i_2+i_3+i_4=n} {\Spab{\la_1+z\la_2|
(q_1+ R_1)(q_1+R_2+z\la_2\W\la_1)(q_1+R_3)|\W\la_2-z\W\la_1}\over
(q_1+ R_1+z \la_2\W\la_1)^2  (q_1+ R_2+z \la_2\W\la_1)^2(q_1+ R_3+z
\la_2\W\la_1)^2} \nn & & \times {A_{i_1 i_2 i_3 i_4,\phi}\over R_1^2
(R_2-R_1)^2 (R_3-R_2)^2 (R_4-R_3)^2}~,~~~\label{IB-exp}\eea
where $R_1=\sum_{j=1}^{i_1} p_j$, $R_2=\sum_{j=1}^{i_1+i_2} p_j$,
$R_3=\sum_{j=1}^{i_1+i_2+i_3} p_j$, $R_4=\sum_{j=1}^{n} p_j$, and
\bea  A_{i_1 i_2 i_3 i_4,\phi} & = & A_{i_1+1}(p_1,...,p_{i_1},
p_{R_{i_1}}) A_{i_2+1} A_{i_3+1} A_{i_4+1}\eea
where $A_{i_k+1}$, $k=2,3,4$ have similar structure like $A_{i_1+1}$
so we have written them briefly. Taking the residue we have
\bea B[I_B] & = & (-ig)^4 \sum_{i_1+i_2+i_3+i_4=n}
 {-1\over \Spab{2|q_1+R_2|1}} {A_{i_1 i_2 i_3 i_4,\phi}\over R_1^2
(R_2-R_1)^2 (R_3-R_2)^2 (R_4-R_4)^2}~.~~~\label{IB-exp}\eea
Having these two boundary contributions (\ref{IA-exp})
and(\ref{IB-exp}) we can finally write down the boundary BCFW
recursion relation as

\bea & & A_{n+2}(q_1^-;p_1,...,p_n;q_2^+) = B[I_A]+B[I_B]\nn
& + & \sum_{i=1,h=\pm}^{n-1}
A_{i+2}(q_1^+(z_i);p_1,...,p_i;q_i^h(z_i)){1\over (q_1+\sum_{j=1}^i
p_j)^2}A_{n-i+2}(-q_i^{-h}(z_i);p_{i+1},...,p_n;q_2^-(z_i))~.
 ~~~~\label{12-+BCFW}\eea

There is one thing we want to remark for this helicity. Our above
analysis is done with the $\Spab{1|2}$-shifting. However, if we use
the $\Spba{1|2}$-shifting, it is easy to see that there is no
boundary contribution. Thus we can calculate same amplitudes using
two different methods: one with boundary contribution and one
without. This will be a strong consistent check for our formula.

\subsection{Explicit calculations for various helicity configurations}

In this subsection we will use the BCFW recursion relations
presented in previous subsection to calculate various amplitudes
with all possible helicity configurations. All results are same as
the one given by Feynman diagrams in Appendix A.

\subsubsection{The helicity $(h_{q_1},h_{q_2})=(+,-)$}

This is the simplest case where boundary contributions are zero.
However, as we have emphasized, the sub-amplitudes used in the
recursion relation will involve other helicity configurations, thus
the knowledge of boundary behavior is essential.

With two scalars, there is  only one possible channel ${\cal
I}=\{q_1^+,1\}$ with $z=\frac{P_{q_11}^2}{\gb{2|P_{q_11}|1}}$ where
we have defined $P_{q_1 i}=q_1+p_1+...+p_i$. Putting this we have
 \bea
 &&  A\Big ( q^+_1;p_1,p_2;q^-_2\Big )
 = A\Big ( \WH q^+_1;p_1;-\WH P^+_{q_11}\Big )\frac{1}{P_{q_11}^2}
     A\Big ( \WH P^-_{q_11};p_2;\WH q^-_1 \Big )=
      (-ig)^2\frac{\tgb{1|P_{q_1 1}|2}}{P_{q_1 1}^2}
 \eea

With four scalars there are three channels and we have
\bea
 &&  A\Big ( q^+_1;p_1,p_2,p_3,p_4;q^-_2\Big )\nonumber\\
 &=& (-ig)^4\sum_{i=1}^3\frac{\tgb{1|(P_{q_11}+z_iq)(P_{q_12}+z_iq)(P_{q_13}+z_iq)|2}}
     {\prod_{\substack{j=1 \\ j \ne i}}^3
     \left(1-\frac{z_i}{z_j}\right)P_{q_11}^2P_{q_12}^2P_{q_13}^2}\nonumber\\
 &&  +(-ig)^2(-i\la)\Big [ \frac{\tgb{1|P_{q_1 1}|2}}{P_{q_1 1}^2}
     +\frac{\tgb{1|P_{q_1 3}|2}}{P_{q_1 1}^2}\Big ]
 \eea
where $z_i$ are poles for these three channels. Using identities
\bea\label{eq:id}
        \sum_{i=1}^n\ \prod_{\substack{j=1 \\ j \ne i}}^n
        \frac{1}{1-\frac{z_i}{z_j}}=1,~~
        \sum_{i=1}^n\ \prod_{\substack{j=1 \\ j \ne i}}^n \frac{z^m}{1-\frac{z_i}{z_j}}=1,
        (m\in \mathbb{Z}, 1\leq m<n),~~
        \sum_{i=1}^n\ \prod_{\substack{j=1 \\ j \ne i}}^n
          \frac{z^n}{1-\frac{z_i}{z_j}}=\prod_{k=1}^nz_k
\eea
the first term can be reduced to
\bea
 (-ig)^4\frac{\tgb{1|P_{q_1 1}P_{q_1 2}P_{q_1 3}|2}}{P_{q_1 1}^2P_{q_1 2}^2P_{q_1 3}^2}
\eea
thus we get the same answer as in Appendix.

With six scalars there are five channels. Using identity
(\ref{eq:id}) we can simplify the expression and get
\bea & & A\Big( q^-_1;p_1,p_2,p_3,p_4;q^+_2\Big)
=(-ig)^6\frac{\tgb{1|P_{q_1 1}P_{q_1 2}P_{q_1 3}P_{q_1 4}P_{q_1
5}|2}}
    {P_{q_1 1}^2P_{q_1 2}^2P_{q_1 3}^2P_{q_1 4}^2P_{q_1 5}^2}\\
&&  +(-ig)^4(-i\la)\Big[ \frac{\tgb{1|P_{q_1 1}P_{q_1 2}P_{q_1
3}|2}}
    {P_{q_1 1}^2P_{q_1 2}^2P_{q_1 3}^2P_{46}^2}
    +\frac{\tgb{1|P_{q_1 1}P_{q_1 2}P_{q_1 5}|2}}{P_{q_1 1}^2P_{q_1 2}^2P_{q_1 5}^2P_{35}^2}
    +\frac{\tgb{1|P_{q_1 1}P_{q_1 4}P_{q_1 5}|2}}{P_{q_1 1}^2P_{q_1 4}^2P_{q_1 5}^2P_{24}^2}
    +\frac{\tgb{1|P_{q_1 3}P_{q_1 4}P_{q_1 5}|2}}{P_{q_1 3}^2P_{q_1 4}^2P_{q_1 5}^2P_{13}^2}\Big ]\nonumber\\
&&  +(-ig)^2(-i\la)^2\Big \{
    \frac{\tgb{1|P_{q_1 3}|2}}{P_{q_13}^2P_{13}^2P_{46}^2}
    +\frac{\tgb{1|P_{q_1 1}|2}}{P_{q_1 1}^2P_{26}^2}[\frac{1}{P_{24}^2}
    +\frac{1}{P_{35}^2}+\frac{1}{P_{46}^2}]
    +\frac{\tgb{1|P_{q_1 5}|2}}{P_{q_15}^2P_{15}^2}[\frac{1}{P_{13}^2}
    +\frac{1}{P_{24}^2}+\frac{1}{P_{35}^2}]\Big \}\nonumber \eea
which is again the right one.

\subsubsection{The helicity $(h_{q_1},h_{q_2})=(-,+)$}

This is the most complicated helicity configuration because there
are two possible boundary contributions $B[I_A]$ and $B[I_B]$ with
one or three fermion propagators respectively. Since for these two
case (A) and (B), there are also corresponding pole contributions,
we can combine pole part and boundary part together. Sometimes this
combination gives simpler expression. Noticing this we move to the
explicit calculations.

With only two scalars, there are only one $B[I_A]$ contribution and
one pole contribution:
\bea
   A\Big ( q^-_1;p_1,p_2;q^+_2\Big )= A\Big
 ( \WH q^-_1;p_1;-\WH P^-_{q_11}\Big )\frac{1}{P_{q_11}^2}
     A\Big ( \WH P^+_{q_11};p_2;\WH q^+_2 \Big )+B[I_A]
 \eea
The boundary term of case (A) is given by
 \bea
 B[I_A] & = & (-ig)^2 { (R_{\cal I}^2+ 2R_{\cal I}\cdot
q_2)\over \Spab{2|R_{\cal I}|1}} {A_{i,\phi}\over R_{\cal I}^2
(q_1+q_2+R_{\cal I})^2} \eea
where $R_{\cal I}=P_{\cal I}+q_1$. Adding them up we have
\bea\label{eq:id3} && -(ig)^2\frac{\gb{\la _1+z_{\cal I}\la
_2|P_{\cal I}
    +z_{\cal I}q|\wtl _2-z_{\cal I}\wtl _1}}
    {P_{\cal I}^2}{A_{i,\phi}\over R_{\cal I}^2
    (q_1+q_2+R_{\cal I})^2}+B[I_A]\nonumber\\
&=&-(ig)^2\frac{\gb{1|P_{\cal I}
    |2}}
    {P_{\cal I}^2}{A_{i,\phi}\over R_{\cal I}^2
    (q_1+q_2+R_{\cal I})^2}
    \eea
Using result (\ref{eq:id3}) and and
$z_1=\frac{P_{q_11}^2}{\gb{2|P_{q_11}|1}}$ we can simplify to
\bea
   A\Big ( q^-_1;p_1,p_2;q^+_2\Big )=
     (-ig)^2\frac{\gb{1|P_{q_1 1}|2}}{P_{q_1 1}^2}
 \eea

With four scalars, both case (A) and case (B) give boundary
contributions. For case (A) there are two  nonzero contributions
with splitting $(1|234)$ and $(123|4)$ (the splitting $(12|34)$ will
give zero). For case (B) there is only one contribution with
splitting $(1|2|3|4)$. For pole part we have three channels ${\cal
I}=\{q_1,p_1\},\{q_1,p_1,p_2\},\{q_1,p_1,p_2,p_3\}$ with location of
poles $z_1,z_2,z_3$ respectively. Putting all together we have
expression
 \bean
 &&  A\Big ( q^-_1;p_1,p_2,p_3,p_4;q^+_2\Big )\nonumber\\
 &=& A\Big ( \WH q^-_1;p_1;-\WH P^-_{q_11}\Big )\frac{1}{P_{q_11}^2}
     A\Big ( \WH P^+_{q_11};p_2,p_3,p_4;\WH q^+_2 \Big )+B[I_A(1|234)]\nonumber\\
 &&  +A\Big ( \WH q^-_1;p_1,p_2;-\WH P^+_{q_11}\Big )\frac{1}{P_{q_12}^2}
     A\Big ( \WH P^-_{q_11};p_3,p_4;\WH q^+_2 \Big )\nonumber\\
 &&  +A\Big ( \WH q^-_1;p_1,p_2,p_3;-\WH P^-_{q_11}\Big )\frac{1}{P_{q_13}^2}
     A\Big ( \WH P^+_{q_11};p_4;\WH q^+_2 \Big )+B[I_A(123|4)]+B[I_B(1|2|3|4)]\nonumber\\
 &=& (-ig)^4\sum_{i=1}^3\frac{\gb{\la_1+z_i\la_2|(P_{q_11}+z_iq)(P_{q_12}+z_iq)(P_{q_13}+z_iq)|\wtl_2-z_i\wtl_1}}
     {\prod_{\substack{j=1 \\ j \ne i}}^3
     \left(1-\frac{z_i}{z_j}\right)P_{q_11}^2P_{q_12}^2P_{q_13}^2}+B[I_B(1|2|3|4)]+B[I_A(1|234)]\nonumber\\
 &&  +(-ig)^2(-i\la)\Big [ \frac{\gb{\la_1+z_1\la_2|P_{q_1 1}|\wtl_2-z_1\wtl_1}}{P_{q_1 1}^2}
     +\frac{\gb{\la_1+z_3\la_2|P_{q_1 3}|\wtl_2-z_3\wtl_1}}{P_{q_1 1}^2}\Big]
     +B[I_A(123|4)]
 \eean
It can be checked that  terms with $z_i^3$ will cancel with the
contribution from $B[I_B(1|2|3|4)]$. Terms with lower order of $z_i$
sum to zero according to the identity in (\ref{eq:id}).
$z_i^3$-independent terms equals $(-ig)^4\frac{\gb{1|P_{q_1 1}P_{q_1
2}P_{q_1 3}|2}}{P_{q_1 1}^2P_{q_1 2}^2P_{q_1 3}^2}$ also by identity
in (\ref{eq:id}). Finally we have
 \bea
 A\Big ( q^-_1;p_1,p_2,p_3,p_4;q^+_2\Big )
=(-ig)^4\frac{\gb{1|P_{q_1 1}P_{q_1 2}P_{q_1 3}|2}}{P_{q_1
1}^2P_{q_1 2}^2P_{q_1 3}^2} +(-ig)^2(-i\la)\Big [ \frac{\gb{1|P_{q_1
1}|2}}{P_{q_1 1}^2}+\frac{\gb{1|P_{q_1 3}|2}}{P_{q_1 3}^2}\Big ]
 \eea

With six scalars, there are five pole channels. For boundary
contributions of case (A), there are three nonzero partitions
$(1|23456)$, $(123|456)$ and $(12345|6)$. For the case (B), there
are four nonzero partitions $(1|2|3|456)$, $(123|4|5|6)$,
$(1|234|5|6)$ and $(1|2|345|6)$. Adding them up and simplifying with
(\ref{eq:id}) we get
\bea && A\Big ( q^-_1;p_1,p_2,p_3,p_4,p_5,p_6;q^+_2\Big )\nonumber \\
&=& (-ig)^6\frac{\gb{1|P_{q_1 1}P_{q_1 2}P_{q_1 3}P_{q_1 4}P_{q_1
5}|2}}
    {P_{q_1 1}^2P_{q_1 2}^2P_{q_1 3}^2P_{q_1 4}^2P_{q_1 5}^2}\nonumber\\
&&  +(-ig)^4(-i\la)\Big[ \frac{\gb{1|P_{q_1 1}P_{q_1 2}P_{q_1 3}|2}}
    {P_{q_1 1}^2P_{q_1 2}^2P_{q_1 3}^2P_{46}^2}
    +\frac{\gb{1|P_{q_1 1}P_{q_1 2}P_{q_1 5}|2}}{P_{q_1 1}^2P_{q_1 2}^2P_{q_1 5}^2P_{35}^2}
    +\frac{\gb{1|P_{q_1 1}P_{q_1 4}P_{q_1 5}|2}}{P_{q_1 1}^2P_{q_1 4}^2P_{q_1 5}^2P_{24}^2}
    +\frac{\gb{1|P_{q_1 3}P_{q_1 4}P_{q_1 5}|2}}{P_{q_1 3}^2P_{q_1 4}^2P_{q_1 5}^2P_{13}^2}\Big ]\nonumber\\
&&  +(-ig)^2(-i\la)^2\Big \{
    \frac{\gb{1|P_{q_1 3}|2}}{P_{q_13}^2P_{13}^2P_{46}^2}
    +\frac{\gb{1|P_{q_1 1}|2}}{P_{q_1 1}^2P_{26}^2}[\frac{1}{P_{24}^2}
    +\frac{1}{P_{35}^2}+\frac{1}{P_{46}^2}]
    +\frac{\gb{1|P_{q_1 5}|2}}{P_{q_15}^2P_{15}^2}[\frac{1}{P_{13}^2}
    +\frac{1}{P_{24}^2}+\frac{1}{P_{35}^2}]\Big \}\eea
which can be checked with result in Appendix A.

\subsubsection{The helicity $(h_{q_1},h_{q_2})=(+,+)$}

In this case, we have $n=odd$ for nonzero results. With $n=1$, there
is no pole contribution, but there is one boundary contribution and
we have
\bea
  A(q_1^+;p;q_2^+)=(-ig)\left[\lambda_1|
  \lambda_2-z\lambda_1\right]=(-ig)\left[1|2\right]
\eea
which can be checked to be right.  With $n=3$ there
 are two pole contributions and one boundary contribution $
  B_{n=3}
  =(-ig)(-i\lambda)\frac{1}{P_{13}^2}\left[1|2\right]$ where $P_{ij}=p_i+p_{i+1}+...
  +p_j$.
Adding up we have
\bea
  && \ A(q_1^+;p_1,p_2,p_3;q_2^+)
       \nonumber \\
  &=&{\widehat{A}}(\wh p_1^+;q_1;-\wh P_{q_1 1}^+)\frac{1}{P_{q_1 1}^2}
                 {\widehat{A}}(\wh P_{q_1 1}^-;q_2,q_3;\wh p_2^+)
                 +{\widehat{A}}(\wh p_1^+;q_1,q_2;-\wh P_{q_1 2}^-)
                 \frac{1}{P_{q_1 2}^2}{\widehat{A}}(\wh P_{q_1 2}^+;q_3;\wh
                 p_2^+)+B_{n=3}
        \nonumber \\
  &=&{(-ig)}^3\left(\left[\widetilde\lambda_1\right|\frac{(P_{q_1 1}-z_1q)
                 (P_{q_1 2}-z_1q)}{P_{q_1 1}^2(P_{q_1 2}^2-z_1P_2q)}
                 \left|\widetilde\lambda_2-z_1\widetilde\lambda_1\right]
                 +\left[\widetilde\lambda_1\right|
                 \frac{(P_{q_1 1}-z_2q)(P_{q_1 2}-z_2q)}{(P_{q_1 1}^2-z_2P_{q_1 1}q)P_{q_1 2}^2}
                 \left|\widetilde\lambda_2-z_2\widetilde\lambda_1\right]\right)
        \nonumber \\
  && +(-ig)(-i\lambda)\frac{1}{P_{13}^2}\left[1|2\right]
        \nonumber \\
  &=&{(-ig)}^3\left[1\right|\frac{{P}_{q_1 1}}{P_{q_1 1}^2}\frac{{P}_{q_1 2}}{P_{q_1 2}^2}\left|2\right]
       +(-ig)(-i\lambda)\frac{1}{P_{13}^2}\left[1|2\right]
\eea

With $n=5$ the boundary part is given by $
  B_{n=5}
  =(-ig){(-i\lambda)}^2\frac{\left[1|2\right]}{P_{15}^2}
  \left(\frac{1}{P_{13}^2}+\frac{1}{P_{24}^2}+\frac{1}{P_{35}^2}\right)
$
Adding the pole part together we have
\bea
  &&\ A(q_1^+;p_1,p_2,p_3,p_4,p_5;q_2^+)  =  B_{n=5}
      \nonumber \\
  && \ +\widehat{A}((\wh q_1^+;p_1;-\wh P_{q_1 1}^+)\frac{1}{P_{q_1 1}^2}
        \widehat{A}((\wh P_{q_1 1}^-;p_2,p_3,p_4,p_5;\wh q_2^+)
       +\widehat{A}((\wh q_1^+;p_1,p_2;-\wh P_{q_1 2}^-)\frac{1}{P_{q_1 2}^2}
       \widehat{A}((\wh P_{q_1 2}^+;p_3,p_4,p_5;\wh q_2^+)
      \nonumber \\
  && \ +\widehat{A}((\wh q_1^+;p_1,p_2,p_3;-\wh P_{q_1 3}^+)\frac{1}{P_{q_1 3}^2}
         \widehat{A}((\wh P_{q_1 3}^-;p_4,p_5;\wh q_2^+)
       +\widehat{A}((\wh q_1^+;p_1,p_2,p_3,p_4;-\wh P_{q_1 4}^-)\frac{1}{P_{q_1 4}^2}
       \widehat{A}((\wh P_{q_1 4}^+;p_5;\wh q_2^+)
 \nn
  &=&(-ig){(-i\lambda)}^2\frac{\left[1|2\right]}{P_{15}^2}\left(\frac{1}{P_{13}^2}+\frac{1}{P_{24}^2}+\frac{1}{P_{35}^2}\right)
         +{(-ig)}^5\left[\widetilde\lambda_1\right|\frac{P_{q_1 1}P_{q_1 2}P_{q_1 3}P_{q_1 4}}
         {P_{q_1 1}^2P_{q_1 2}^2P_{q_1 3}^2P_{q_1 4}^2}\left|\widetilde\lambda_2\right]
      \nonumber \\
  && \ +{(-ig)}^3(-i\lambda)\left(\left[\widetilde\lambda_1\right|
         \frac{P_{q_1 3}P_{q_1 4}}{P_{q_1 3}^2P_{q_1 4}^2}\left|\widetilde\lambda_2\right]
         +\left[\widetilde\lambda_1\right|\frac{P_{q_1 1}P_{q_1 4}}{P_{q_1 1}^2P_{q_1 4}^2}\left|\widetilde\lambda_2\right]
         +\left[\widetilde\lambda_1\right|\frac{P_{q_1 1}P_{q_1 2}}{P_{q_1 1}^2P_{q_1 2}^2}\left|\widetilde\lambda_2\right]\right)
\eea
which is the right one.

\subsubsection{The helicity $(h_{q_1},h_{q_2})=(-,-)$}

This is, in fact, similar to previous one so we will be briefly. For
$n=1$, similar to the case $(+,+)$, there is one boundary
contribution and the result is $B_{n=1}=(-ig)\langle1|2\rangle$. For
$n=3$, we calculate one boundary contribution
$B_{n=3}=(-ig)(-i\lambda)\frac{1}{Q_{13}^2}\left\langle1|2\right\rangle$
and two pole contributions as following:

\bea
  && \ A(q_1^-;p_1,p_2,p_3;q_2^-)
       \nonumber \\
  &=&B_{n=3}
       +{\widehat{A}}(p_1^-;q_1;-P_{q_1 1}^-)\frac{1}{P_{q_1 1}^2}{\widehat{A}}(P_{q_1 1}^+;q_2,q_3;p_2^-)
                 +{\widehat{A}}(p_1^-;q_1,q_2;-P_{q_1 2}^+)\frac{1}{P_{q_1 2}^2}{\widehat{A}}(P_{q_1 2}^-;q_3;p_2^-)
        \nonumber \\
  &=&(-ig)(-i\lambda)\frac{1}{P_{13}^2}\left\langle1|2\right\rangle
        \nonumber \\
  && \ +{(-ig)}^3\left(\left\langle\lambda_1+z_1\lambda_2\right|\frac{(P_{q_1 1}-z_1q)(P_{q_1 2}-z_1q)}
                 {P_{q_1 1}^2(P_{q_1 2}^2-z_1P_{q_1 2}q)}\left|\lambda_2\right\rangle
                 +\left\langle\lambda_1+z_2\lambda_2\right|\frac{(P_{q_1 1}-z_2q)(P_{q_1 2}-z_2q)}
                 {(P_{q_1 1}^2-z_2P_{q_1 1}q)P_{q_1 2}^2}\left|\lambda_2\right\rangle\right)
        \nonumber \\
  &=&{(-ig)}^3\left\langle1\right|\frac{P_1}{P_1^2}\frac{{P}_2}{P_2^2}\left|2\right\rangle
\eea
With $n=5$, the boundary part is given by $
  B[\langle5\rangle]
  =(-ig){(-i\lambda)}^2\frac{\left\langle1|2\right\rangle}{P_{15}^2}
  \left(\frac{1}{P_{13}^2}+\frac{1}{P_{24}^2}+\frac{1}{P_{35}^2}\right)$,
  thus we have
\bea
  &&\ A(q_1^-;p_1,p_2,p_3,p_4,p_5;q_2^-)
      =B_{n=5}
      \nonumber \\
  && \ +\widehat{A}((q_1^-;p_1;-P_{q_1 1}^-)\frac{1}{P_{q_1 1}^2}
        \widehat{A}((P_{q_1 1}^+;p_2,p_3,p_4,p_5;q_2^-)
       +\widehat{A}((q_1^-;p_1,p_2;-P_{q_1 2}^+)\frac{1}{P_{q_1 2}^2}
       \widehat{A}((P_{q_1 2}^-;p_3,p_4,p_5;q_2^-)
      \nonumber \\
  && \ +\widehat{A}((q_1^-;p_1,p_2,p_3;-P_{q_1 3}^-)\frac{1}{P_{q_1 3}^2}
         \widehat{A}((P_{q_1 3}^+;p_4,p_5;q_2^-)
       +\widehat{A}((q_1^-;p_1,p_2,p_3,p_4;-P_{q_1 4}^+)\frac{1}{P_{q_1 4}^2}
       \widehat{A}((P_{q_1 4}^-;p_5;q_2^-)
      \nonumber \\
  &=&(-ig){(-i\lambda)}^2\frac{\left\langle1|2\right\rangle}{P_{15}^2}\left(
         \frac{1}{P_{13}^2}+\frac{1}{P_{24}^2}+\frac{1}{P_{35}^2}\right)
         +{(-ig)}^5\left\langle\lambda_1\right|\frac{P_{q_1 1}P_{q_1 2}P_{q_1 3}P_{q_1 4}}
         {P_{q_1 1}^2P_{q_1 2}^2P_{q_1 3}^2P_{q_1 4}^2}\left|\lambda_2\right\rangle
      \nonumber \\
  && \ +{(-ig)}^3(-i\lambda)\left(\left\langle\lambda_1\right|
         \frac{P_{q_1 3}P_{q_1 4}}{P_{q_1 3}^2P_{q_1 4}^2}\left|\lambda_2\right\rangle
         +\left\langle\lambda_1\right|\frac{P_{q_1 1}P_{q_1 4}}{P_{q_1 1}^2P_{q_1 4}^2}
         \left|\lambda_2\right\rangle
         +\left\langle\lambda_1\right|\frac{P_{q_1 1}P_{q_1 2}}{P_{q_1 1}^2P_{q_1 2}^2}
         \left|\lambda_2\right\rangle\right)
\eea
is exactly what we get from Feynman diagrams.

\section{Conclusion}

In this paper, we have analyzed the on-shell constructibility more
carefully. We showed that for some theories, although there is no
any deformation which has vanishing boundary contribution, the
boundary contributions can still be identified and calculated
on-shell recursively. With the knowledge of boundary behavior we can
write down the generalized BCFW recursion relation.

Our examples in this paper is simple in this sense that all boundary
contributions can be directly analyzed by just a few Feynman
diagrams. There are other examples where above direct analysis is
not so straightforward, for example, the gauge theory with
deformation $\Spab{1^-|2^+}$. It will be interesting to find other
methods to deal with these more complicated situations. Having the
knowledge of boundary behavior we can have better idea for the
off-shell and on-shell constructibility. Then we may have better
understanding of S-matrix theory using, for example, the technique
presented in \cite{Paolo:2007}.

There are many questions we can ask for ourselves. For example, the
rational part of one-loop amplitudes can be calculated by BCFW
recursion relation too (see, for example, \cite{Bern:2005hs}-
\cite{Berger:2006ci}). It is found that sometimes there is nonzero
boundary contribution. Using our new understanding, it maybe useful
to recheck this problem.

\section*{Acknowledgments}

This work is funded by Qiu-Shi funding from Zhejiang University and
Chinese NSF funding under contraction No.10875104. We would like to
thank Prof. Luo for useful discussions and Prof. Britto, Dr. Wu for
reading the draft.

\appendix

\section{Amplitudes from Feynman diagrams}

In this appendix, we will calculate various amplitudes using Feynman
diagrams directly to compare with results from boundary BCFW
recursion relation. For simplicity we will focus on the ordered
results.

\subsection{Amplitude of Pure Scalar Field }

The Feynman rule for this theory is very simple: there is only one
vertex with four scalar lines and coupling constant $-i\la$. Using
this we get following results.

\subsubsection{6-Point Amplitude}

\begin{figure}[!h]
  \centering
  \includegraphics[viewport=160 653 305 715,clip]{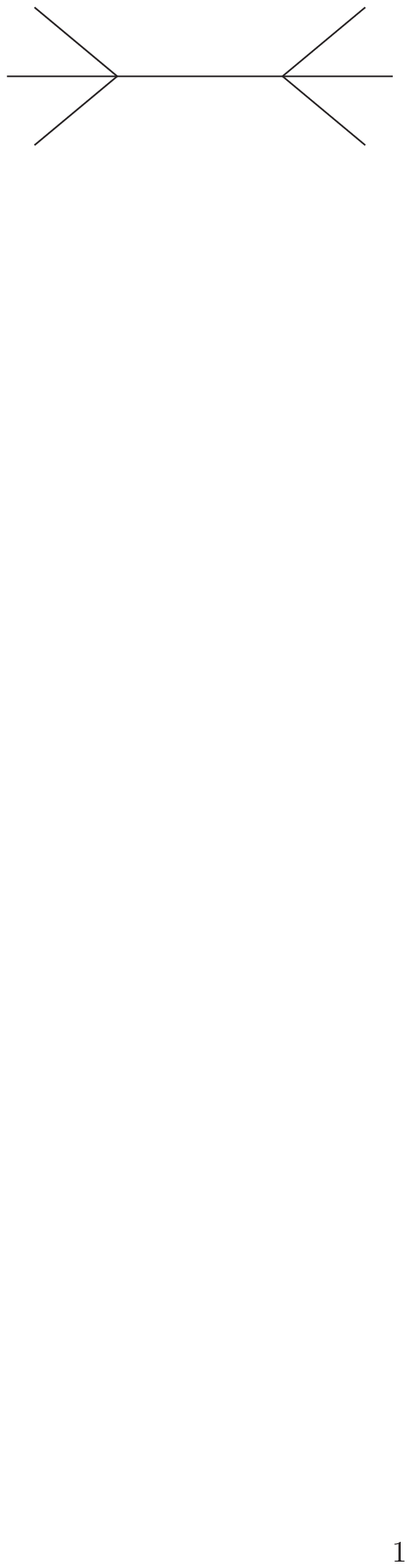}
  \caption{A 6-point Feynman diagram}\label{Fig-6-point}
\end{figure}

The typical Feynman diagrams for the 6-point amplitude are given in
Figure \ref{Fig-6-point}.
There are six of them by cyclic ordering. Each one is given by
$(-i\la)^2\frac{1}{P_{i(i+1)(i+2)}^2}$ where $P_{ijk}=p_i+p_j+p_k$.
Adding them up we have
\bea
 A_6^{FD}(1,\dots,6)=(-i\lambda)^2\left(\frac{1}{P_{123}^2}+\frac{1}{P_{126}^2}
 +\frac{1}{P_{156}^2}\right)~,~~~~\label{6-FD}
\eea
where we have identified $i+6\equiv i$.

\subsubsection{8-Point Amplitude}

\begin{figure}[!h]
  \centering
  \includegraphics[viewport=140 635 485 709,clip]{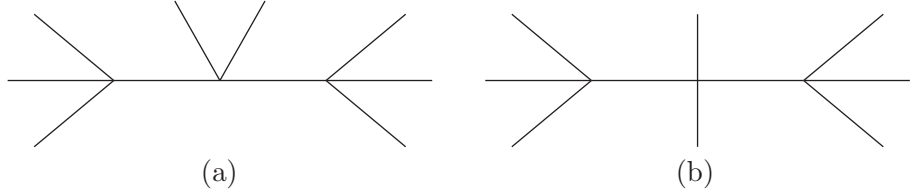}
  \caption{Three kinds of 8-point Feynman diagrams}\label{Fig-8-pint}
\end{figure}

The diagrams for 8-point amplitude have two different topologies as
given in Figure \ref{Fig-8-pint}.
The expression of the first kind of diagrams has the form
      \bea
                (-i\lambda)^3\frac{1}{P_{i(i+1)(i+2)}^2
                  P_{(i+5)(i+6)(i+7)}^2}~
      \eea
with eight cyclic orderings.
 The second kind of Feynman diagram has the form
      \bea
                (-i\lambda)^3\frac{1}{P_{i(i+1)(i+2)}^2
                  P_{(i+3)(i+4)(i+5)}^2}
      \eea
with only four cyclic orderings, since  it is obvious that this
figure is same with shifting $i\to i+4$. Adding them up we have
      \bea
        A_8^{FD}(1,\dots,8)&=&A_8^{FD(a)}+A_8^{FD(b)}
                    \nonumber \\
                &=&(-i\lambda)^3\sum_{\sigma\in\mathbb{Z}_8}\left(\frac{1}{P_{\sigma(1)\sigma(2)\sigma(3)}^2
                  P_{\sigma(6)\sigma(7)\sigma(8)}^2}+\frac{1}{2P_{\sigma(1)\sigma(2)\sigma(3)}^2P_{
                  \sigma(5)\sigma(6)\sigma(7)}^2}\right)~.~~~\label{8-FD}
      \eea
%

\subsubsection{10-Point Amplitude}

The possible seven topologies  of the diagrams are given in
following:
\begin{figure}[!h]
\label{Fig_8} \centering
  \includegraphics[viewport=137 649 364 715
                   ,scale=0.9
                   ,clip]{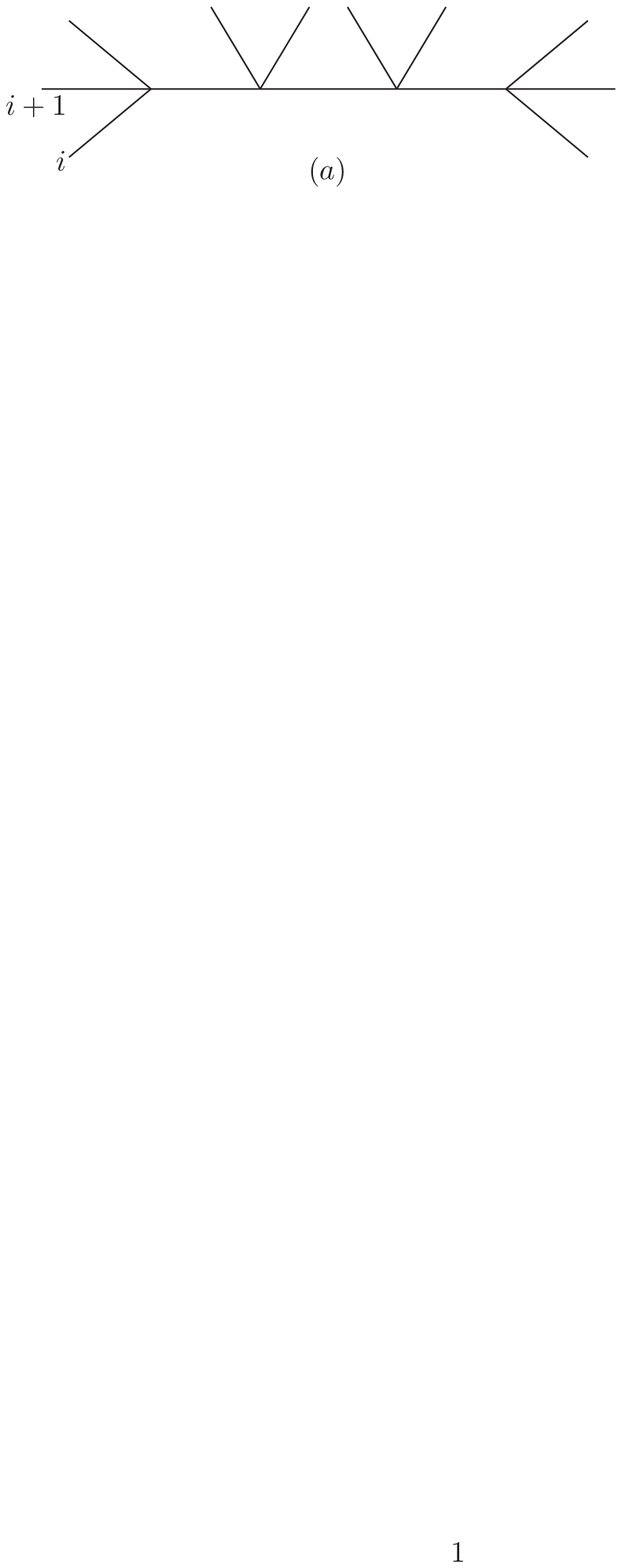}
  \includegraphics[viewport=137 649 364 715
                   ,scale=0.9
                   ,clip]{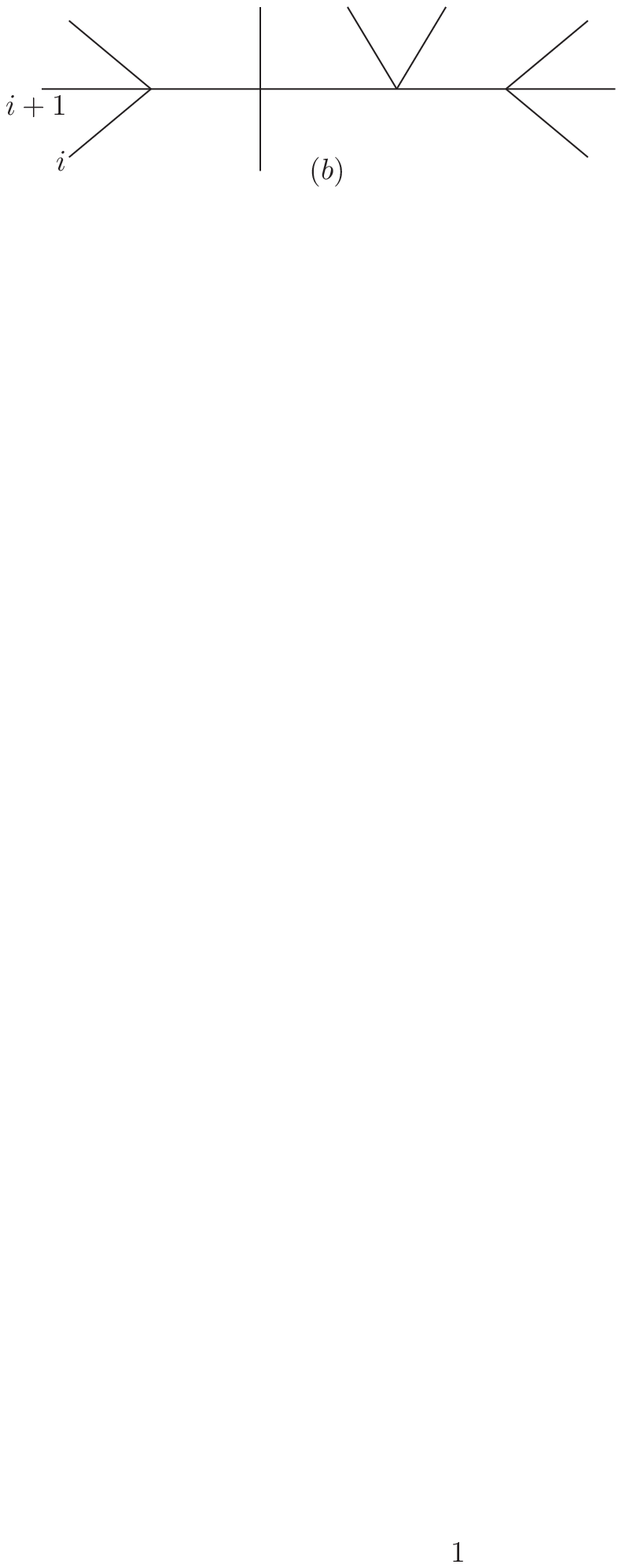}
\end{figure}
\begin{figure}[!h]
\centering
  \includegraphics[viewport=137 649 364 715
                   ,scale=0.9
                   ,clip]{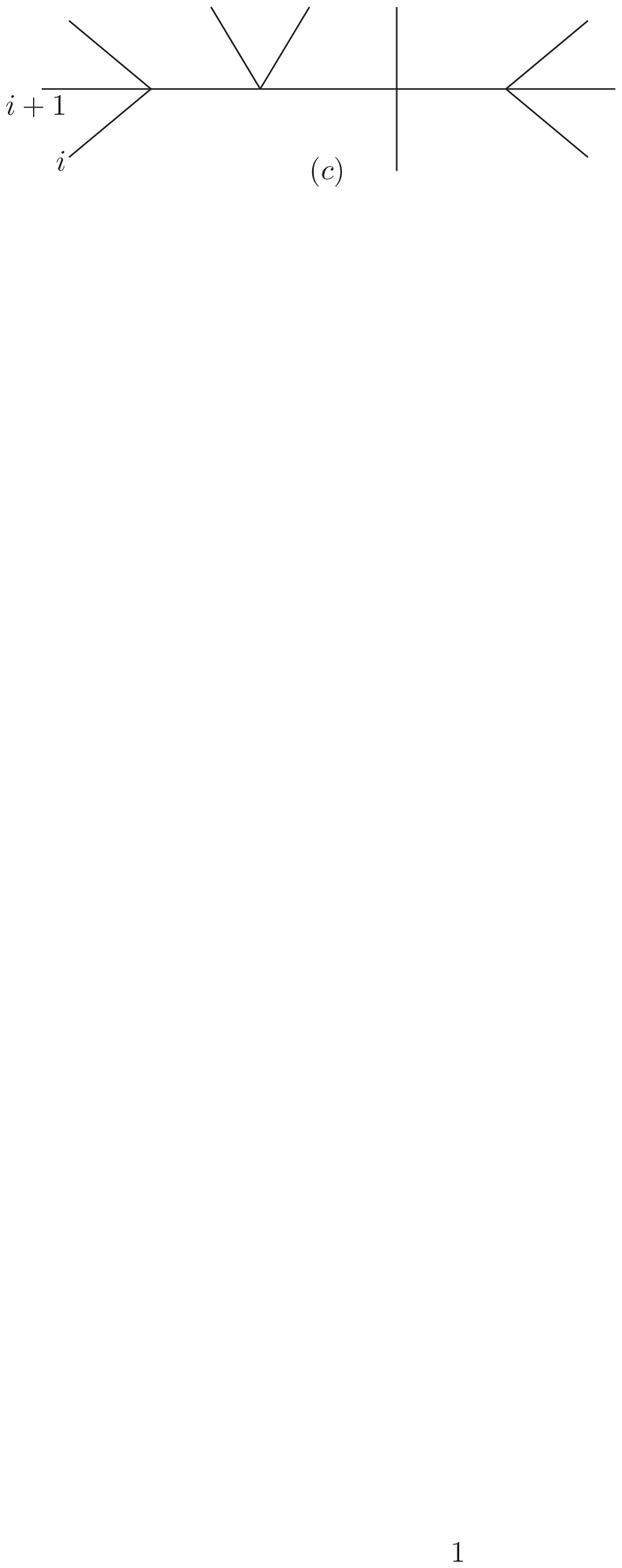}
  \includegraphics[viewport=137 649 364 715
                   ,scale=0.9
                   ,clip]{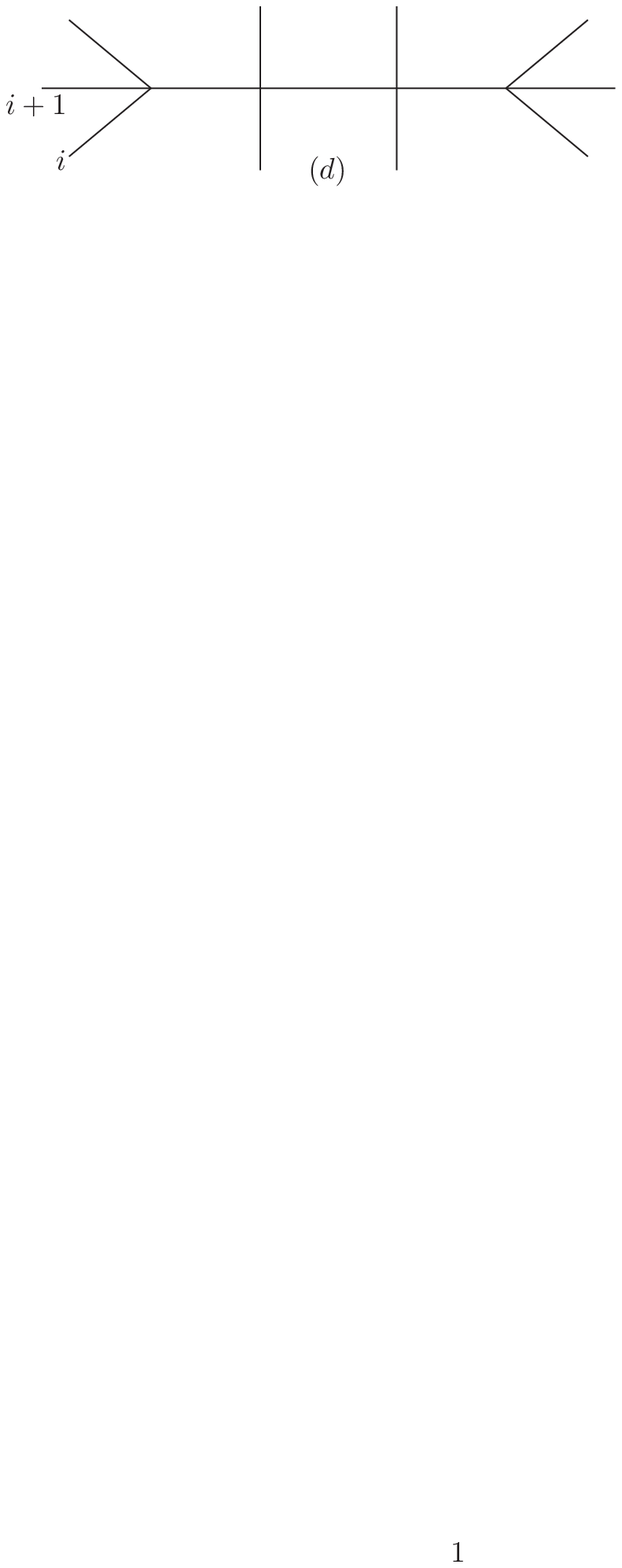}
\end{figure}
\begin{figure}[!h]
\centering
  \includegraphics[viewport=137 649 364 715
                   ,scale=0.9
                   ,clip]{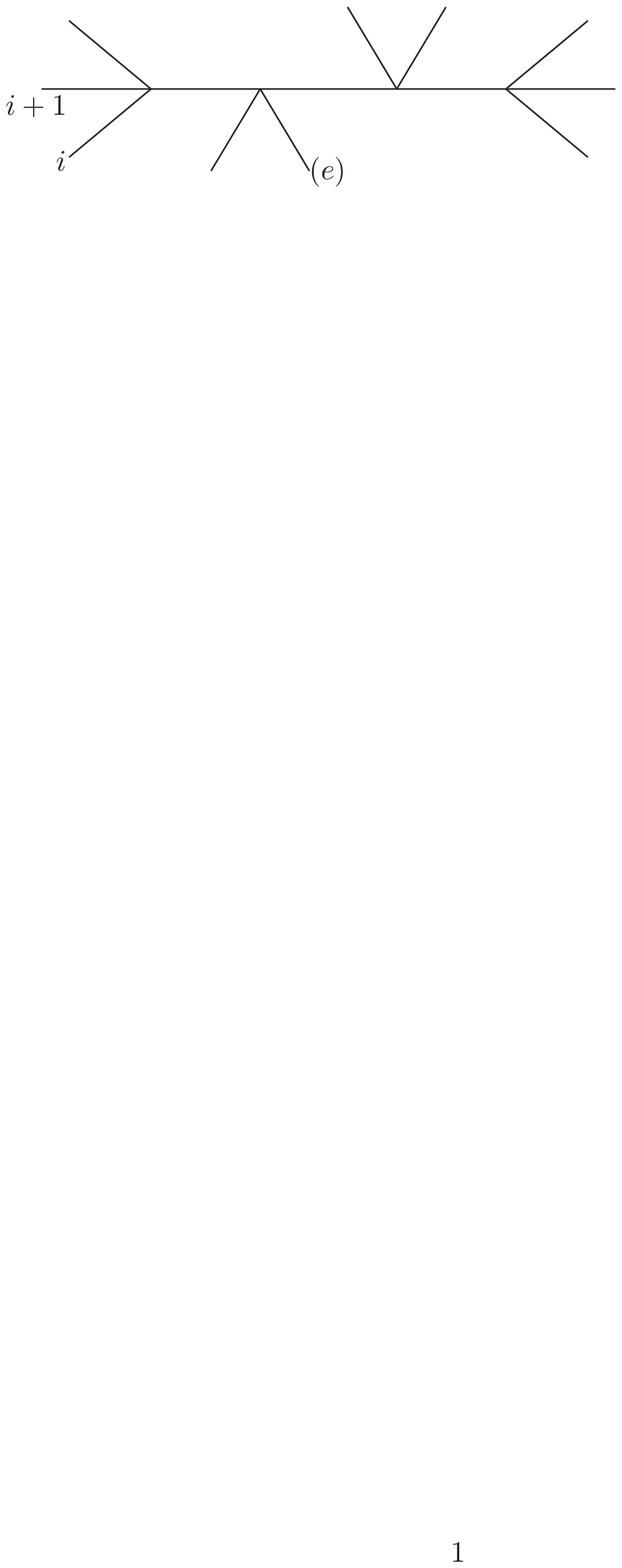}
  \includegraphics[viewport=137 649 364 715
                   ,scale=0.9
                   ,clip]{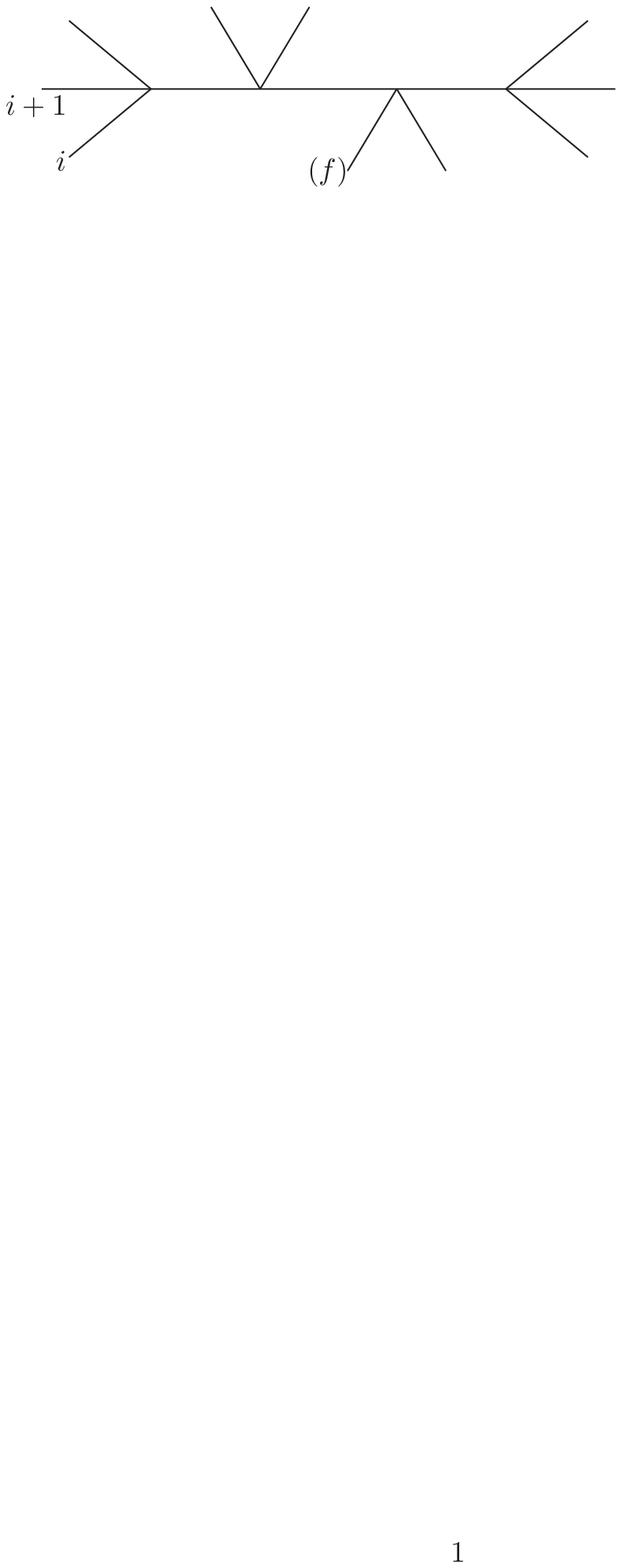}
\end{figure}
\begin{figure}[!h]
  \centering
  \includegraphics[viewport=137 528 284 700
                   ,scale=0.9
                   ,clip]{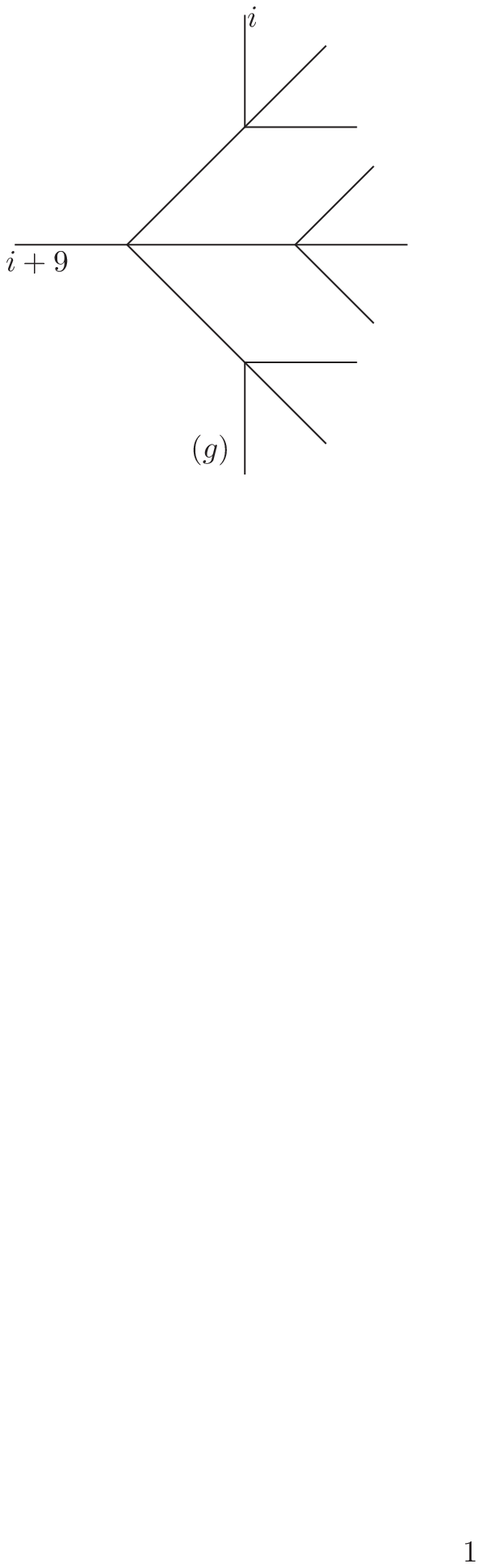}
\end{figure}
and the corresponding expressions are given as
 \bea \label{Eqt:1}
  {\cal{Q}}_{(a)}&=&\frac{1}{P^2_{i(i+1)(i+2)}P^2_{i(i+1)(i+2)(i+3)(i+4)}
                P^2_{(i+7)(i+8)(i+9)}}
\\
  {\cal{Q}}_{(b)}&=&\frac{1}{P^2_{i(i+1)(i+2)}P^2_{i(i+1)(i+2)(i+3)(i+4)}
                P^2_{(i+6)(i+7)(i+8)}}
\\
  {\cal{Q}}_{(c)}&=&\frac{1}{P^2_{i(i+1)(i+2)}P^2_{(i-1)i(i+1)(i+2)(i+3)}
                P^2_{(i+6)(i+7)(i+8)}}
\\
  {\cal{Q}}_{(d)}&=&\frac{1}{P^2_{i(i+1)(i+2)}P^2_{(i-2)(i-1)i(i+1)(i+2)}
                P^2_{(i+5)(i+6)(i+7)}}
\\
  {\cal{Q}}_{(e)}&=&\frac{1}{P^2_{i(i+1)(i+2)}P^2_{i(i+1)(i+2)(i+3)(i+4)}
                P^2_{(i+5)(i+6)(i+7)}}
\\
  {\cal{Q}}_{(f)}&=&\frac{1}{P^2_{i(i+1)(i+2)}P^2_{(i-1)i(i+1)(i+2)(i+3)}
                P^2_{(i+5)(i+6)(i+7)}}
\\
  {\cal{Q}}_{(g)}&=&\frac{1}{P^2_{i(i+1)(i+2)}P^2_{(i+3)(i+4)(i+5)}
                P^2_{(i+6)(i+7)(i+8)}}
\eea

Among these seven topologies,  three of them, i.e., $(d)$, $(e)$,
$(f)$, are intrinsic symmetric under $\sigma:[i]\mapsto[i+5]$, while
remaining four are full $Z_{10}$ ordering.  Thus the final answer is
given by
 \bea
        && \  A_{10}^{FD}(1,2,\dots,10) \nonumber \\
        &=&{(-i\lambda)}^4\sum_{\sigma\in{\mathbb{Z}}_{10}}\Biggl[
                \frac{1}{P^2_{\sigma(1)\sigma(2)\sigma(3)}P^2_{\sigma(1)\sigma(2)\sigma(3)\sigma(4)\sigma(5)}
                P^2_{\sigma(8)\sigma(9)\sigma(10)}}+\frac{1}{P^2_{\sigma(1)\sigma(2)\sigma(3)}
                P^2_{\sigma(1)\sigma(2)\sigma(3)\sigma(4)\sigma(5)}P^2_{\sigma(7)\sigma(8)\sigma(9)}}
                    \nonumber \\
        &&      +\frac{1}{2P^2_{\sigma(1)\sigma(2)\sigma(3)}P^2_{\sigma{(10)}\sigma(1)\sigma(2)\sigma(3)\sigma(4)}
                P^2_{\sigma(6)\sigma(7)\sigma(8)}}+\frac{1}{P^2_{\sigma(1)\sigma(2)\sigma(3)}
                P^2_{\sigma{(10)}\sigma(1)\sigma(2)\sigma(3)\sigma(4)}P^2_{\sigma(7)\sigma(8)\sigma(9)}}
                    \nonumber \\
        &&      +\frac{1}{2P^2_{\sigma(1)\sigma(2)\sigma(3)}P^2_{\sigma(1)\sigma(2)\sigma(3)\sigma(4)\sigma(5)}
                P^2_{\sigma(6)\sigma(7)\sigma(8)}}
                 +\frac{1}{2P^2_{\sigma(1)\sigma(2)\sigma(3)}P^2_{\sigma(1)\sigma(2)\sigma(3)\sigma(9)\sigma(10)}
                P^2_{\sigma(6)\sigma(7)\sigma(8)}}\nonumber\\
        &&      +\frac{1}{P^2_{\sigma(1)\sigma(2)\sigma(3)}P^2_{\sigma(4)\sigma(5)\sigma(6)}
                P^2_{\sigma(7)\sigma(8)\sigma(9)}}\Biggr]~~~\label{10-point}
      \eea

\subsection{Amplitude of two fermions and $n$ scalars}

\begin{figure}[htb]
   \centering
   \includegraphics[viewport=140 388 327 704
                    ,clip]{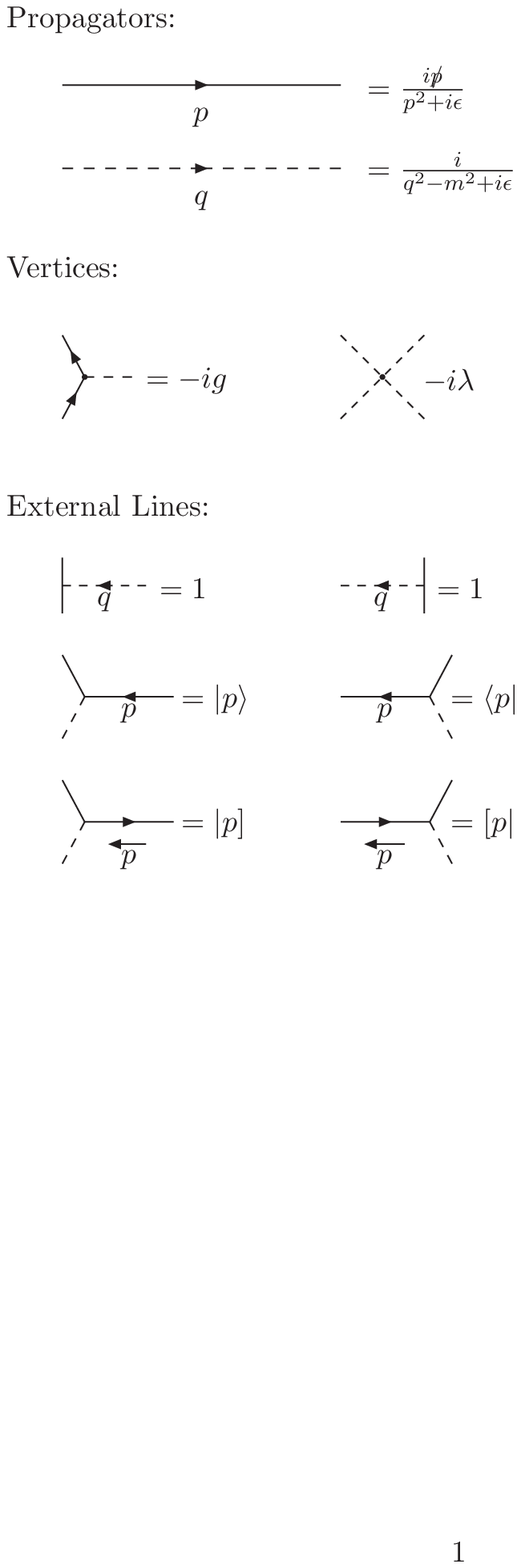}
   \caption{Feynman rules for fermion-scalar field}\label{F-S-Rule}
 \end{figure}

The Feynman rules for ordered scalar QCD is given by Figure
\ref{F-S-Rule}. Using this we can calculate various amplitudes
$A_{2,n}(q_1;p_1,...,p_n; q_2)$ where $q_1, q_2$ for two fermions
and $p_i$ for scalars. For these amplitudes, we  need to notice that
since the scalar part has only $\la\phi^4$ vertex, $A_{2,n}$ with
helicities $(h_{q_1},h_{q_2})=(+,+)/(-,-)$  is not zero only when
$n=odd$ while with $(h_{q_1},h_{q_2})=(+,-)/(-,+)$ it is not zero
only when $n=even$. It is also important to notice that when we
write $\ket{1},\ket{2}$ they mean $\ket{\la_{q_1}},\ket{\la_{q_2}}$.
%

\subsubsection{Amplitude of $A(q_1^+, q_2^+; p_1,...,p_n)$}

For this case we have $n=odd$. For this case we have $n=odd$. With
$n=1$ we have
\bea
   A(q_1^+;p;q_2^+)=(-ig)\left[1 | 2\right]
\eea
With $n=3$ we have
\bea
  A(q_1^+;p_1,p_2,p_3;q_2^+)={(-ig)}^3\frac{\left[1\right|{P}_{p_11}{P}_{p_12}\left|2\right]}
   {P_{p_11}^2P_{p_12}^2}
   +(-ig)(-i\lambda)\frac{1}{P_{13}^2}\left[1|2\right]
\eea
where we have defined $P_{q_1 i}=q_1+p_1+...+p_i$ and
$P_{ij}=p_i+p_{i+1}+...+p_j$ With $n=5$ we have
\bea
  &&\ A(q_1^+;p_1,p_2,p_3,p_4,p_5;q_2^+)
      \nonumber \\
  &=&{(-ig)}^5\frac{\left[1\right|P_{p_11}P_{p_12}P_{p_13}P_{p_14}\left|2\right]}
               {P_{p_11}^2P_{p_12}^2P_{p_13}^2P_{p_14}^2}
               +{(-ig)}^3(-i\lambda)\left(\frac{\left[1\right|P_{p_13}P_{p_14}\left|2\right]}
               {P_{p_13}^2P_{p_14}^2}+\left[1\right|\frac{P_{p_11}P_{p_14}\left|2\right]}
               {P_{p_11}^2P_{p_14}^2}+\frac{\left[1\right|P_{p_11}P_{p_12}\left|2\right]}
               {P_{p_11}^2P_{p_12}^2}\right)
      \nonumber \\
  && \ +(-ig){(-i\lambda)}^2\frac{\left[1|2\right]}
      {P_{15}^2}\left(\frac{1}{P_{13}^2}+\frac{1}{P_{24}^2}+\frac{1}{P_{35}^2}\right)
\eea
%

\subsubsection{Amplitude of $A(q_1^-, q_2^-; p_1,...,p_n)$}

With $n=1$ we have
\bea A(q_1^-;p;q_2^-)=(-ig)\langle1|2\rangle\eea
With $n=3$ we have
\bea
    A^{FD}(q_1^-;p_1,p_2,p_3;q_2^-)={(-ig)}^3\frac{\left\langle 1\right|
    {P}_{p_11}{P}_{p_12}\left|2\right\rangle}{P_{p_11}^2P_{p_12}^2}
    +(-ig)(-i\lambda)\frac{1}{P_{13}^2}\left\langle1|2\right\rangle
\eea
With $n=5$ we have
\bea
  &&\ A^{FD}(q_1^-;p_1,p_2,p_3,p_4,p_5;q_2^-)
      \nonumber \\
  &=&{(-ig)}^5\frac{\left\langle1\right|P_{p_11}P_{p_12}P_{p_13}P_{p_14}\left|2\right\rangle}
            {P_{p_11}^2P_{p_12}^2P_{p_13}^2P_{p_14}^2}+{(-ig)}^3(-i\lambda)\left(
            \frac{\left\langle1\right|P_{p_13}P_{p_14}\left|2\right\rangle}{P_{p_13}^2P_{p_14}^2}
            +\frac{\left\langle1\right|P_{p_11}P_{p_14}\left|2\right\rangle}{P_{p_11}^2P_{p_14}^2}
            +\frac{\left\langle1\right|P_{p_11}P_{p_12}\left|2\right\rangle}{P_{p_11}^2P_{p_12}^2}\right)
      \nonumber \\
  && \ +(-ig){(-i\lambda)}^2\frac{\left\langle1|2\right\rangle}
          {P_{15}^2}\left(\frac{1}{P_{13}^2}+\frac{1}{P_{24}^2}
          +\frac{1}{P_{35}^2}\right)
\eea
%

\subsubsection{Amplitude of $A(q_1^+, q_2^-; p_1,...,p_n)$}

In this case we need to have even number of $n$. With $n=2$ we have
\bea A\Big ( q^+_1;p_1,p_2;q^-_2\Big ) =(-ig)^2\frac{\tgb{1|P_{q_1
1}|2}}{P_{q_1 1}^2}~. ~~~~~\label{Fermion+-2}\eea
where because the color ordering we have defined $P_{q_1
i}=q_1+p_1+p_2+...+p_i$.  With $n=4$ we have
\bea A\Big ( q^+_1;p_1,p_2,p_3,p_4;q^-_2\Big )
=(-ig)^4\frac{\tgb{1|P_{q_1 1}P_{q_1 2}P_{q_1 3}|2}}{P_{q_1
1}^2P_{q_1 2}^2P_{q_1 3}^2} +(-ig)^2(-i\la)\Big [
\frac{\tgb{1|P_{q_1 1}|2}}{P_{q_1 1}^2}+\frac{\tgb{1|P_{q_1
3}|2}}{P_{q_1 1}^2}\Big ]~. ~~~~~\label{Fermion+-4} \eea
With $n=6$ we have
\bea
&&  A\Big ( q^+_1;p_1,p_2,p_3,p_4,p_5,p_6;q^-_2\Big )\nonumber\\
&=& (-ig)^6\frac{\tgb{1|P_{q_1 1}P_{q_1 2}P_{q_1 3}P_{q_1 4}P_{q_1
5}|2}}
    {P_{q_1 1}^2P_{q_1 2}^2P_{q_1 3}^2P_{q_1 4}^2P_{q_1 5}^2} ~~~~~\label{Fermion+-6}\\
&&  +(-ig)^4(-i\la)\Big[ \frac{\tgb{1|P_{q_1 1}P_{q_1 2}P_{q_1
3}|2}}
    {P_{q_1 1}^2P_{q_1 2}^2P_{q_1 3}^2P_{46}^2}
    +\frac{\tgb{1|P_{q_1 1}P_{q_1 2}P_{q_1 5}|2}}{P_{q_1 1}^2P_{q_1 2}^2P_{q_1 5}^2P_{35}^2}
    +\frac{\tgb{1|P_{q_1 1}P_{q_1 4}P_{q_1 5}|2}}{P_{q_1 1}^2P_{q_1 4}^2P_{q_1 5}^2P_{24}^2}
    +\frac{\tgb{1|P_{q_1 3}P_{q_1 4}P_{q_1 5}|2}}{P_{q_1 3}^2P_{q_1 4}^2P_{q_1 5}^2P_{13}^2}\Big ]\nonumber\\
&&  +(-ig)^2(-i\la)^2\Big \{
    \frac{\tgb{1|P_{q_1 3}|2}}{P_{q_13}^2P_{13}^2P_{46}^2}
    +\frac{\tgb{1|P_{q_1 1}|2}}{P_{q_1 1}^2P_{26}^2}[\frac{1}{P_{24}^2}
    +\frac{1}{P_{35}^2}+\frac{1}{P_{46}^2}]
    +\frac{\tgb{1|P_{q_1 5}|2}}{P_{q_15}^2P_{15}^2}[\frac{1}{P_{13}^2}
    +\frac{1}{P_{24}^2}+\frac{1}{P_{35}^2}]\Big \}\nonumber~.
\eea
\\

\subsubsection{Amplitude of $A(q_1^-, q_2^+; p_1,...,p_n)$}

With $n=2$ we have
 \bea A\Big ( q^-_1;p_1,p_2;q^+_2\Big )
=(-ig)^2\frac{\gb{1|P_{q_1 1}|2}}{P_{q_1 1}^2}~.
~~~~~\label{Fermion-+2} \eea
With $n=4$ we have
 \bea A\Big ( q^-_1;p_1,p_2,p_3,p_4;q^+_2\Big )
=(-ig)^4\frac{\gb{1|P_{q_1 1}P_{q_1 2}P_{q_1 3}|2}}{P_{q_1
1}^2P_{q_1 2}^2P_{q_1 3}^2} +(-ig)^2(-i\la)\Big [ \frac{\gb{1|P_{q_1
1}|2}}{P_{q_1 1}^2}+\frac{\gb{1|P_{q_1 3}|2}}{P_{q_1 1}^2}\Big ] ~.
~~~~~\label{Fermion-+4}\eea
With $n=6$ we have
\bea
&&  A\Big ( q^-_1;p_1,p_2,p_3,p_4,p_5,p_6;q^+_2\Big )\nonumber\\
&=&(-ig)^6\frac{\gb{1|P_{q_1 1}P_{q_1 2}P_{q_1 3}P_{q_1 4}P_{q_1
5}|2}} {P_{q_1 1}^2P_{q_1 2}^2P_{q_1 3}^2P_{q_1 4}^2P_{q_1 5}^2}
~~~~~\label{Fermion-+6}\\
&&  +(-ig)^4(-i\la)\Big[ \frac{\gb{1|P_{q_1 1}P_{q_1 2}P_{q_1 3}|2}}
    {P_{q_1 1}^2P_{q_1 2}^2P_{q_1 3}^2P_{46}^2}
    +\frac{\gb{1|P_{q_1 1}P_{q_1 2}P_{q_1 5}|2}}{P_{q_1 1}^2P_{q_1 2}^2P_{q_1 5}^2P_{35}^2}
    +\frac{\gb{1|P_{q_1 1}P_{q_1 4}P_{q_1 5}|2}}{P_{q_1 1}^2P_{q_1 4}^2P_{q_1 5}^2P_{24}^2}
    +\frac{\gb{1|P_{q_1 3}P_{q_1 4}P_{q_1 5}|2}}{P_{q_1 3}^2P_{q_1 4}^2P_{q_1 5}^2P_{13}^2}\Big ]\nonumber\\
&&  +(-ig)^2(-i\la)^2\Big \{
    \frac{\gb{1|P_{q_1 3}|2}}{P_{q_13}^2P_{13}^2P_{46}^2}
    +\frac{\gb{1|P_{q_1 1}|2}}{P_{q_1 1}^2P_{26}^2}[\frac{1}{P_{24}^2}
    +\frac{1}{P_{35}^2}+\frac{1}{P_{46}^2}]
    +\frac{\gb{1|P_{q_1 5}|2}}{P_{q_15}^2P_{15}^2}[\frac{1}{P_{13}^2}
    +\frac{1}{P_{24}^2}+\frac{1}{P_{35}^2}]\Big \}~. \nonumber
\eea
\\

\section{Amplitudes With More External Fermions}

In section four, we have focused on the case with only two fermions.
In this Appendix, we will discuss the general case with multiple
pair of fermions. For amplitudes with $2l$ fermions and $n$ scalars,
there are $l$ fermion lines, which are connected to each other by
scalar lines. The amplitudes will be the form
\bea A=\sum_i {\cal S}_i \prod_{j=1}^l {\cal Q}_{ij} \eea
where ${\cal S}_i$ are scalar parts and each ${\cal Q}_{i,j}$ is
following form
\bea {\cal Q}_{i,j}(q_i^-, q_j^+; R_1,...,R_m)\sim
i^m{\Spab{\la_i|R_1|R_2|...|R_m|\W\la_j}\over R_1^2 R_2^2 ....
R_m^2} \eea
where depending on the helicities of $i,j$, we may need to change
$\la\to \W\la$.

Now we choose two fermions, for example, $q_1, q_2$ to make the
$\Spab{1|2}$-deformation. There are two categories of Feynman
diagrams: (A) two fermions $q_1, q_2$ are connected by same fermion
line; (B) two fermions $q_1, q_2$ locate at different fermion line
and are connected through scalar propagators.

For the category (A), the boundary behavior is exactly the same one
as we have discussed in section four with only two fermions. Thus we
can write down similar  boundary contributions and add them to the
boundary BCFW recursion relation.

For the category (B) things are different. First  there is at least
one scalar propagator connecting fermion lines and having ${1\over
z}$ dependence. Second, when there are two nearby fermion
propagators along same fermion line, because
\bea & & {\Spaa{\a|(R_1+z\la_2 \W\la_1)(R_2+z\la_2 \W\la_1)|\b}\over
(R_1+z\la_2 \W\la_1)^2(R_2+z\la_2 \W\la_1)^2}\nn & = &
{\Spaa{\a|(R_1+z\la_2 \W\la_1)R_2|\b}\over (R_1+z\la_2
\W\la_1)^2(z\la_2 \W\la_1)^2}+{\Spaa{\a|R_1(z\la_2 \W\la_1)|\b}\over
(R_1+z\la_2 \W\la_1)^2(R_2+z\la_2 \W\la_1)^2}~,\eea
we have another ${1\over z}$ dependence instead of naive
$z^0$-dependence. Using above two observations,
we can  discuss case by case:\\
\begin{itemize}

\item (B-1) {\bf Helicity $(h_{q_1},h_{q_2})=(+,+)$:}\\

For this case, along the line with $z$-dependence (this line will be
constituted by fermion propagators and scalar propagators), we have
one $z$ from external wave-function of $q_2$, i.e.,
$\bket{\W\la_2-z\W\la_1}$, ${1\over z^s}$ from $s\geq 1$ scalar
propagators and $f$ fermion propagators with naive $z^0$-dependence.
However, when there are $m$ pair nearby propagators  as discussed
above, we have another  ${1\over z^m}$-dependence. Putting all
together, we found that to have nonzero boundary contributions, we
need to satisfy following conditions: (a) there is only one scalar
propagator depending on $z$; (b) there is no any nearby
$z$-depending fermion propagator pair.

The condition (a) implies that the line connecting $q_1,q_2$
involves only two fermion lines. Furthermore, condition (b) tells us
that there is at most one fermion propagator along each fermion line
(remembering that we have two fermion lines here). Using
$(R+z\la_2\W\la_1)\bket{\W\la_2-z\W\la_1}$ is only order of $z$ and
$[\W\la_1|(R+z\la_2\W\la_1|=[\W\la_1|R|$, we see that no any
$z$-depending fermion propagator can be nearby the external particle
$q_1,q_2$, thus the only boundary contribution is the one without
any fermionic propagator depending on $z$ as shown by Figure
\ref{Fig:special}. The Figure can also be represented as
\bea C\Spbb{\W\la_1|\a}{1\over (R+z\W\la_2\la_1)^2}
\Spbb{\b|\W\la_2-z\W\la_1} \eea
where $\a,\b,C$ are $z$-independent part from other components of
diagrams.

      \begin{figure}[htb]
  \centering
  \includegraphics[viewport=152 613 260 700,clip]{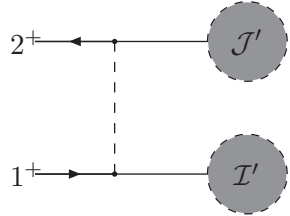}
  \caption{special case\label{Fig:special}}
\end{figure}

The helicity $(-,-)$ will be similar and we will not discuss it further. \\

\item (B-2) {\bf Helicity $(h_{q_1},h_{q_2})=(+,-)$:}\\

For this case, along the line with $z$-dependence, we have  ${1\over
z^s}$ from $s\geq 1$ scalar propagators and $f$ fermion propagators
with naive $z^0$-dependence. Thus there is no boundary
contribution.\\

\item (B-3) {\bf Helicity $(h_{q_1},h_{q_2})=(-,+)$:}\\

For this case, along the line with $z$-dependence, we have  $z^2$
from external wave-functions $\bket{\W\la_2-z\W\la}$ and
$\ket{\la_1+z\la_2}$, ${1\over z^s}$ from $s\geq 1$ scalar
propagators and $f$ fermion propagators with naive $z^0$-dependence.
This is the most complicated case with many possibilities, so we
list them one-by one.

With only one scalar propagator (i.e., $s=1$), we have several
diagrams which can be  represented as following:
\bea (I-1): & ~~~& C\Spab{\la_1+z\la_2|\a} {1\over
(Q+z\W\la_2\la_1)^2} \Spbb{\b|\W\la_2-z\W\la_1}\sim z^1\nn
(I-2): & ~~~& C\Spab{\la_1+z\la_2|(R_1+z\la_2\W\la_1)|\a} {1\over
(Q+z\W\la_2\la_1)^2(R_1+z\la_2\W\la_1)^2}
\Spbb{\b|\W\la_2-z\W\la_1}\sim z^0\nn
(I-3): & ~~~& C\Spab{\la_1+z\la_2|\a} {1\over
(Q+z\W\la_2\la_1)^2(R_1+z\la_2\W\la_1)^2}
\Spbb{\b|(R_1+z\la_2\W\la_1)|\W\la_2-z\W\la_1}\sim z^0\nn
(I-4): & ~~~&
C{\Spab{\la_1+z\la_2|(R_1+z\la_2\W\la_1)(R_2+z\la_2\W\la_1)|\a}
\over (R_1+z\la_2\W\la_1)^2(R_2+z\la_2\W\la_1)^2(Q+z\W\la_2\la_1)^2}
\Spbb{\b|\W\la_2-z\W\la_1}\sim z^0\nn
(I-5): & ~~~& C\Spab{\la_1+z\la_2|\a}
{\Spbb{\b|(R_1+z\la_2\W\la_1)(R_1+z\la_2\W\la_1)|\W\la_2-z\W\la_1}\over
(R_1+z\la_2\W\la_1)^2(R_1+z\la_2\W\la_1)^2(Q+z\W\la_2\la_1)^2}\sim
z^0\nn \eea
The case (I-1) tells us that there is no any $z$-depending fermion
propagator. Case (I-2) and (I-3) represent the diagrams with one
$z$-depending fermion propagator nearby external particles $q_1,
q_2$ respectively. Case (I-4) and (I-5) represent the diagrams with
two $z$-depending fermion propagators nearby external particles
$q_1, q_2$ respectively. It is worth to notice that we have not
included the case where we have two $z$-depending fermion
propagators and each external $q_i$ has one propagator nearby, since
this one has vanished boundary contribution.

 With two scalar propagators, there
are again two situations we need to consider. The first one is that
there are only two fermion lines involved. The expression for this
one is
\bea (II-1): & ~~~& C\Spab{\la_1+z\la_2|\a} {1\over
(Q_1+z\W\la_2\la_1)^2(Q_2+z\W\la_2\la_1)^2}
\Spbb{\b|\W\la_2-z\W\la_1}~.\eea
The second one is that there are three fermion lines involved. For
this one, there are two different representations. The first one is
\bea C\Spab{\la_1+z\la_2|\a} {1\over
(Q_1+z\W\la_2\la_1)^2}\Spbb{\gamma|\delta}{1\over
(Q_2+z\W\la_2\la_1)^2} \Spbb{\b|\W\la_2-z\W\la_1}~,\eea
where there is no fermion propagator in middle fermion line
depending on $z$. However, since we have only triple-vertex $\O\psi
\phi \psi$, this case can not be realized. The second one is
\bea (II-2): C\Spab{\la_1+z\la_2|\a} {1\over
(Q_1+z\W\la_2\la_1)^2}{\Spba{\gamma|(R+z\la_2\W\la)|\delta}\over
(R+z\la_2\W\la)^2}{1\over (Q_2+z\W\la_2\la_1)^2}
\Spbb{\b|\W\la_2-z\W\la_1}~\eea
where there is one fermion propagator in middle fermion line
depending on $z$.

Adding all seven cases together, we get total boundary contribution
when $q_1, q_2$ are not in same fermion line.

\end{itemize}
%


\end{document}